# SCIENCE AND TECHNOLOGY OF *BOREXINO*: A REAL TIME DETECTOR FOR LOW ENERGY SOLAR NEUTRINOS

## Borexino Collaboration

G. Alimonti[k], C. Arpesella[a], H. Back[b], M. Balata[a], T. Beau[n], G. Bellini[k&], J. Benziger[q], S. Bonetti[k], A. Brigatti[k], B. Caccianiga[k], L. Cadonati[r], F. Calaprice[r], G. Cecchet[o], M. Chen[i], A. DeBari[o], E. DeHaas[r], H. de Kerret[n], O. Donghi[a], M. Deutsch[d], F. Elisei[p], A. Etenko[l], F. von Feilitzsch[f], R. Fernholz[r], R. Ford[a], B. Freudiger[h], A. Garagiola[k], C. Galbiati[r], F. Gatti[g], S. Gazzana[a], M. Giammarchi[k], D. Giugni[k], A. Golubchikov[k], A. Goretti[k], C. Grieb[f], C. Hagner[f], T. Hagner[f], W. Hampel[h], E. Harding[r], F. Hartmann[k], R. von Hentig[f], H. Hess[f], G. Heusser[h], A. Ianni[r], P. Inzani[k], S. Kidner[r], J. Kiko[h], T. Kirsten[h], G. Korga[k*], G. Korschinek[f], D. Kryn[n], V. Lagomarsino[g], P. LaMarche[#], M. Laubenstein[a], F. Loeser[r], P. Lombardi[k], S. Magni[k], S. Malvezzi[k], J. Maneira[k***], I. Manno[c], G. Manuzio[g], F. Masetti[p], U. Mazzucato[p], E. Meroni[k], P. Musico[g], H. Neder[h], M. Neff[f], S. Nisi[a], L. Oberauer[f], M. Obolensky[n], M. Pallavicini[g], L. Papp[k*], L. Perasso[k], A. Pocar[r], R. Raghavan[m], G. Ranucci[k], W. Rau[a], A. Razeto[g], E. Resconi[g], T. Riedel[f], A. Sabelnikov[l], P. Saggese[k], C. Salvo[g], R. Scardaoni[a], S. Schoenert[f***], K. Schuhbeck[f], H. Seidel[f], T. Shutt[r], H. Simgen[h], A. Sonnenschein[r], O. Smirnov[e], A. Sotnikov[e], M. Skorokhvatov[l], S. Sukhotin[l], R. Tartaglia[a], G. Testera[g], R. Vogelaar[b], S. Vitale[g], M. Wojcik[j], O. Zaimidoroga[e], Y. Zakharov[h+].

a) Gran Sasso National Laboratories (LNGS), Assergi, Italy
b) Virginia Polytechnic Institute and State University, Blacksburg VA, USA
c) Research Institute for Particle and Nuclear Physics, Budapest, Hungary
d) Massachusetts Institute of Technology, Cambridge MA, USA
e) Joint Institute for Nuclear Research, Dubna, Russia
f) Technical University Munich, Garching, Germany
g) Physics Department of the University and INFN, Genoa, Italy
h) Max Planck Institute for Nuclear Physics, Heidelberg, Germany
i) Queen's University, Kingston, Canada
j) Jagellonian University, Krakow, Poland
k) Physics Department of the University and INFN, Milan, Italy
l) Kurchatov Institute, Moscow, Russia
m) Bell Laboratories, Lucent Technologies, Murray Hill NJ, USA
n) College de France, Paris, France
o) Physics Department of the University and INFN, Pavia, Italy
p) Chemistry Department of the University and INFN, Perugia, Italy
q) Department of Chemical Engineering, Princeton University, Princeton NJ, USA
r) Department of Physics, Princeton University, Princeton NJ, USA

& Spokesperson
# Project manager
* On leave from (c)
** On leave from: Department of Physics, University of Lisbon, Lisbon, Portugal
*** Now at (h)
+ Permanent address: Institute of Nuclear Research, Moscow, Russia

**ABSTRACT**

*Borexino, a real-time device for low energy neutrino spectroscopy is nearing completion of construction in the underground laboratories at Gran Sasso, Italy (LNGS). The experiment's goal is the direct measurement of the flux of $^7$Be solar neutrinos of all flavors via neutrino-electron scattering in an ultra-pure scintillation liquid. Seeded by a series of innovations which were brought to fruition by large scale operation of a 4-ton test detector at LNGS, a new technology has been developed for Borexino. It enables sub-MeV solar neutrino spectroscopy for the first time. This paper describes the design of Borexino, the various facilities essential to its operation, its spectroscopic and background suppression capabilities and a prognosis of the impact of its results towards resolving the solar neutrino problem. Borexino will also address several other frontier questions in particle physics, astrophysics and geophysics.*

## 1. The New Solar Neutrino Problem

In 1968, Ray Davis pioneered an experiment in the Homestake mine (USA) to look directly into the center of the Sun and to probe the thermonuclear fusion reactions that generate solar energy by observing neutrinos (ν) from these reactions[1]. Not only did he accomplish this goal but he also discovered a significant and unexplained shortfall of the solar ν flux compared to astrophysical predictions. Since then, the science and technology of solar ν detection[2,3] as well as the sophistication of models of the solar interior[4,5] and their predictions of ν fluxes have advanced remarkably. The first generation detectors Kamiokande[6] (based on ν-e scattering) and Gallex[7] and Sage[8] (based on $\nu_e$ capture in $^{71}$Ga) that followed the Homestake detector[9] (based on $\nu_e$ capture in $^{37}$Cl) have extended observations over the full spectrum of solar ν's. The flux deficits persist over the entire spectral range and still remain a puzzle. On the other hand, advances in helioseismology leave little room for substantial departures from the Standard Solar Model (SSM) and its predictions of ν fluxes.[5] Consequently the basic thrust of these data - the new solar ν problem - is now focused more sharply on the ν itself and on its non-standard properties, such as non-zero mass, flavor mixing[10], magnetic moment, decay, and others.

A non-standard ν implies physics beyond the Standard Model of elementary particles. For this reason, the second generation solar neutrino experiments are now at the cutting edge of particle physics: Superkamiokande (SK)[11], the Sudbury Neutrino Observatory (SNO)[12], and Borexino. Further, the focus is shifting to the low energy part of the solar neutrino spectrum <1 MeV. In real time, this range is accessible exclusively to Borexino. The Borexino detector is now nearing completion of construction. In particular, it will explore the $^7$Be neutrinos.

The paradigm shift in this field was generated from the incompatibility between the results of Gallex/Sage on the one hand and Kamiokande/SK on the other. These experiments brought new sensitivities for solar ν detection. The gallium experiments set a low energy threshold to observe the integrated ν flux from the proton-proton (pp) reactions, from $^7$Be-decay, and from other reactions in the Sun. Kamiokande/SK have yielded real-time data on the high energy ν flux from a single solar ν source, the decay of $^8$B. The signal observed in Ga, ≈50% of the expected magnitude[13], is practically exhausted by the ν flux from the p+p reactions, a 'must-see' flux nearly independent of solar models *if* neutrinos are mass-less particles as described by the standard model of electro-weak interactions. On this basis, the Ga signals would imply a near total absence of $^7$Be in the Sun. The compelling consequence is a near absence of $^8$B as well, since that can be generated only from the reaction $^7$Be+p→$^8$B. In stark contrast, a substantial flux from decaying $^8$B (even if with a deficit of ≈50%) is observed by Kamiokande/SK. The signal rates at low and high energies are thus internally inconsistent with the sequential logic of energy production in the Sun, with or without invoking modification of the SSM.[14,15,16] This is referred to as the $^7$Be-$^8$B problem.

The validity of the SSM is strongly supported by helioseismological observations.[5] The most likely explanation for the missing ν flux lies in non-standard ν physics and, in particular, in non-zero ν masses that enable ν flavor conversion. The main observable consequence is an *energy dependence* of the flux deficits. If this were observed experimentally, it would unambiguously prove neutrino flavor oscillations. Indeed, several physical scenarios of flavor conversion fit all presently available experimental data and still leave the solar models intact. With these developments, a sharpened solar ν-problem sets well-defined directives to the next generation experiments. The key questions are: What is the ν flux at the Earth from $^7$Be in the Sun? and: How can one demonstrate solar ν flavor conversion directly?

Borexino[17] is designed to measure the $^7$Be-ν induced interaction rate and to answer the central questions phrased above. Early laboratory research on radio-purities of scintillation liquids[18] gave initial hints that the background from natural radioactivity at low energies may be sufficiently controllable to make real-time spectroscopy of sub-MeV solar ν's feasible. Explicit demonstration of that possibility had then to come from the 4-ton test detector CTF (Counting Test Facility).

The plan of this paper is as follows: In Sec. 2, we sketch the physical foundations for solar ν emission, flavor conversion, the likelihood ranges for non-standard ν parameters, basic aspects of ν-e$^-$ scattering, and the ν signal detection process in Borexino. As the first real-time experiment for low energy solar ν spectroscopy, Borexino is based on a new technology, the foundations of which are briefly covered in Sec. 3. Next we describe the Borexino detector and several ancillary facilities as well as the operational details of the detector (Sec. 4). The response of Borexino to various non-standard ν scenarios is then treated (Sec. 5). We conclude with remarks on the outlook for the impact of Borexino on critical topics in non-solar ν science in particular (Sec. 6) and in non-standard ν phenomenology in general (Sec. 7).

## 2. Physical Foundations

### 2.1 The Solar Neutrino Spectrum

The basis of solar energy production is the fusion of 4 protons to form an α-particle, a process that releases 26.73 MeV of energy. The 'PP chain' of reactions that achieves this end (Fig.1) is initiated by the 'pp' reaction p + p → d + e$^+$ + ν$_e$. The pp and its variant 'pep' reaction, p + e$^-$ + p → d + ν$_e$ are driven by the weak interaction and control the slow rate of hydrogen burning in the Sun to last altogether > 9 Gyr. The next steps to the termination to α-particles occur via 3 branches, PPI (86%), PPII (14%) and PPIII ($1.6 \times 10^{-4}$). Neutrinos are emitted in each of these branches. The ν fluxes from the pp and pep reactions in PPI are considered 'standard candle' fluxes since, as initiating reactions, they are relatively independent of the rest of the chain in a quasi-static solar interior and thus, they do not strongly depend on details of the SSM. The fluxes are basically determined by the observed solar luminosity and by the mean energy release in the $4p \rightarrow \alpha$ conversion. Contrary to this, the $\nu_e$ fluxes from the electron-capture decay of $^7$Be in PPII and especially from the $\beta^+$-decay of $^8$B in PPIII depend strongly on the details of the solar model, particularly, the central temperature in the Sun.[19] The CNO cycle, while less important than the PP cycle, still adds non-negligible contributions of $\nu$ fluxes from the $\beta^+$-decays of $^{13}$N, $^{15}$O and $^{17}$F.

The Sun thus emits a rich spectrum of neutrinos whose measurement offers a complete and detailed probe of the PP chain. The most abundant component is the pp-$\nu$ continuum at (0-0.42) MeV. The next feature is the mono-energetic line at 0.862 MeV from $^7$Be (which also produces a weak line at 0.384 MeV under the pp-continuum) followed by another line at 1.44 MeV from the 'pep' reaction. Underlying these features are weak continua from the CNO reactions (0-1.7) MeV. This low energy part of the spectrum covers the dominant PPI and PPII links. The decay of $^8$B in PPIII contributes the smallest flux, $\approx10^{-4}$ of the pp flux, in a continuum up to $\approx14$ MeV. Finally, $\nu$'s with higher energies up to $\approx20$ MeV may be emitted from the poorly characterized $^3$He + p ('hep') reaction. It is clear from Fig.1 that the key link in the PP chain is PPII, a link accessible only via $^7$Be-$\nu$'s. Thus, Borexino is a key *probe of the roles of astrophysics and $\nu$ physics* in the solar $\nu$ problem.

The SSM fluxes $\varphi_\nu$ from the various $\nu_e$ sources are given in Table 1 (see Fig. 2 for the solar $\nu$ spectrum[20]). Uncertainties in these predictions depend on the stage of the reaction in the PP-chain and on the associated nuclear cross-sections. The (1$\sigma$) error estimates of the theoretical fluxes are $\delta\varphi_\nu$(pp)$\approx1\%$, $\delta\varphi_\nu$($^7$Be)$\approx9\%$ and $\delta\varphi_\nu$($^8$B)$\approx20\%$. The relatively large $\delta\varphi_\nu$($^8$B) arises from the acute dependence on the solar temperature (~T$^{20}$) and from the low precision ($\approx20\%$) of the measured cross-section of the production reaction $^7$Be(p,$\gamma$)$^8$B.

## 2.2 Neutrino Flavor Conversion

A direct consequence of a non-degenerate mass spectrum of neutrinos and flavor mixing is flavor conversion.[21] During transport in matter or vacuum, solar $\nu$'s that were inherently emitted as pure $\nu_e$ can be converted spontaneously to other flavors, $\nu_\mu$ or $\nu_\tau$. The latter are either not detected at all (as in charged current based detectors), or observed with reduced (factor $\approx5$) cross-sections (as in neutral current based detection). Flavor conversion can thus create solar $\nu$ signal deficits in a natural way. In a two flavor model with a 2×2 matrix that transforms a doublet of mass eigenstates $\nu_1$ and $\nu_2$ with masses $m_1$ and $m_2$ to flavor states $\nu_e$ and $\nu_x$ via the mixing angle $\theta$, the survival probability of the e-flavor can be written in terms of $\Delta m^2 = m_1^2 - m_2^2$ and $\sin^2(2\theta)$. Patterns of e-flavor survival vs. $\nu$ energy are shown in Figs. 3 and 4 for two distinct physical scenarios. In vacuum, conversion manifests itself as $\nu$ flavor oscillation with a wave length (oscillation length) $\lambda = 4\pi E/\Delta m^2$. Fig. 3 shows the interesting regime when $\lambda \approx R$, the Earth-Sun baseline - the 'just-so' oscillations.[22]

**Table 1**: Predicted ν fluxes arriving at the Earth and their theoretical uncertainties[5] for the various solar ν sources.

| ν Source | Energy [MeV] | ν Flux [$10^6$ cm$^{-2}$ s$^{-1}$] |
|---|---|---|
| p-p | ≤ 0.42 | 59 400 ± 600 |
| pep | 1.44 | 139 ± 1.4 |
| hep | ≤ 18.8 | ≈0.0021 |
| $^7$Be | 0.86 (90%), 0.38 (10%) | 4 800 ± 430 |
| $^8$B | ≤ 15 | $5.15\ ^{+0.98}_{-0.72}$ |
| $^{13}$N | ≤ 1.20 | $605\ ^{+115}_{-79}$ |
| $^{15}$O | ≤ 1.70 | $532\ ^{+117}_{-80}$ |

**Figure 1**: The nuclear reaction sequence of the pp-chain inside the Sun

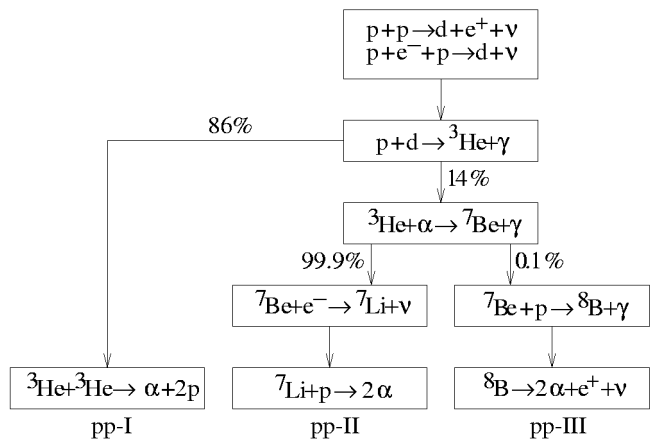

**Figure 2**: The solar neutrino spectrum, from Ref.[20]

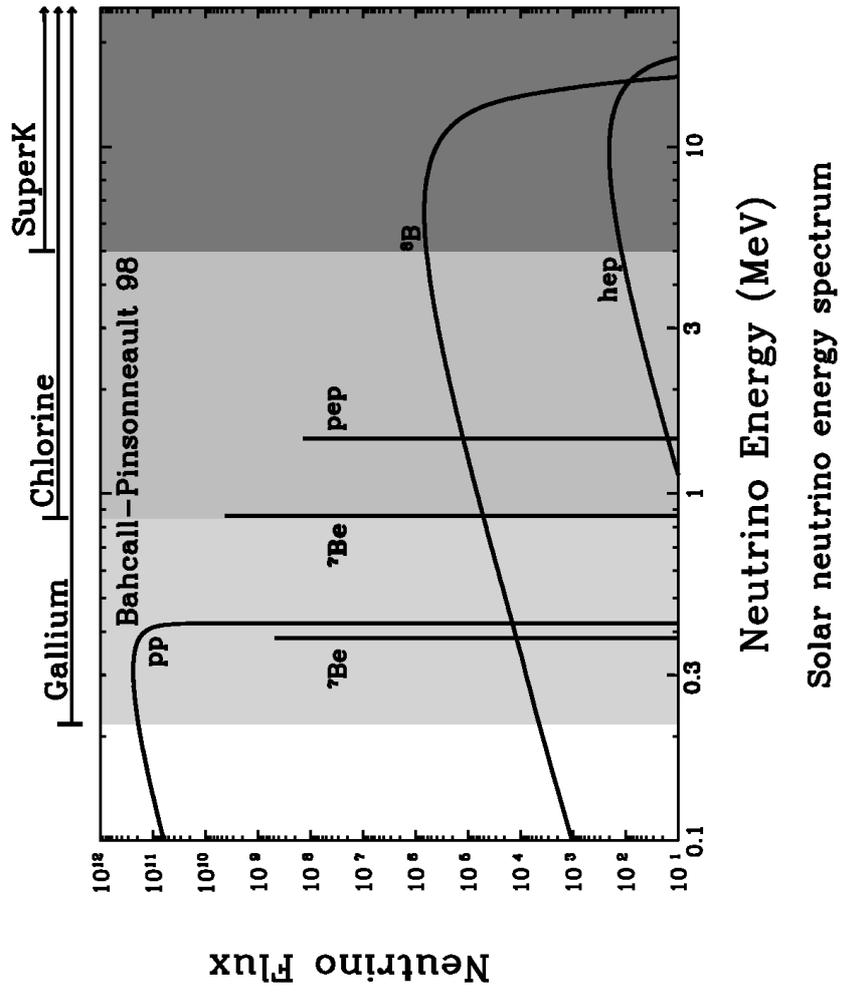

Conversion can also occur as a resonant process in solar matter (the MSW effect).[23] This flavor conversion is energy dependent, especially pronounced at *lower* energies (see Figure 4). The effect offers a convincing explanation for the apparent paradox of the $^7$Be- and $^8$B-$\nu$ signals. Figure 4 shows the survival probabilities for electron neutrinos after flavor conversion for three distinct islands of $(\Delta m^2, \theta)$ that happen to offer solutions that are consistent with all experimental solar neutrino data acquired so far ('SMA', 'LMA', 'LOW').[24]

The basic observable of flavor conversion of solar $\nu$'s is the e-flavor survival probability, measured by a flux 'disappearance' relative to the SSM values for $\varphi_\nu(^7$Be,$^8$B). 'Appearance effects' are also possible as direct proofs of conversion. The energy dependence creates characteristic distortions of the standard weak-interaction spectral shapes of $\nu$ continua, in particular that from $^8$B.

Time variations of the signal are observable in certain scenarios. In vacuum, the phase of the flavor oscillation changes slowly as the Earth-Sun baseline varies in the eccentric Earth orbit, but nevertheless this produces large changes in the flavor survival of a mono-energetic line <1 MeV, such as that from $^7$Be. This results in a *seasonal signal variation*[25] (in addition to the small [≈7%] variation expected just from the $R^{-2}$ effect). Variability even on a *day/night basis* is possible (especially for the $^7$Be line flux) because the converted flavor arriving at the Earth can be reconverted to e-flavor in passing through the Earth matter, thus enhancing the night time signal[26] (see Fig. 4). All these effects on the $^7$Be-$\nu$ line and on the low energy part of the $^8$B $\nu$ spectrum are observable in Borexino (Sec. 5).

Matter and vacuum conversion occur in distinct regions of the $\Delta m^2$-$\sin^2(2\theta)$ map. Vacuum oscillations require large mixing [$\sin^2(2\theta) \approx 1$] that sets the oscillation amplitudes; and tiny mass differences, $\Delta m^2 \approx 10^{-11}$ to $10^{-9}$ that result in experimentally interesting oscillation periods. Matter conversion dominates for $\Delta m^2$ from $10^{-8}$ to $10^{-4}$(eV/c$^2$)$^2$ even for small mixing, with $\sin^2(2\theta)$ ranging from very small mixing angles ($\approx 10^{-3}$) to the maximum value of $\approx 1$. Solar $\nu$ experiments thus probe large parameter spaces in key regimes not accessible to any other approach.

Global analysis of the data from all the present experiments already constrains the $\nu$ parameters to a few islands in the $\Delta m^2$-$\sin^2(2\theta)$ maps[27] for the MSW effect (Fig. 5) and for vacuum oscillations (Fig. 6). The e-flavor survival probability plots in Figures 3 and 4 refer to just these islands in the $\nu$ parameter space.

## 2.3 Neutrino Electron Scattering

Solar neutrino detection in Borexino is based on $\nu$-e$^-$ scattering using the well-known technique of liquid scintillation (LS) spectroscopy. Scintillation photons generated in a large mass of aromatic organic liquid are detected. They yield a measure of the energy and of the spatial positions of the events in the detector. Photon emission in the LS is isotropic, thus, the inherent directionality of the $\nu$-e scattering process cannot be utilized for deriving the directionality of the $\nu$ source.

**Figure 3**: Flavor survival vs. energy of neutrinos in vacuum oscillations for $\Delta m^2 = 6.5 \times 10^{-11} (eV/c^2)^2$ and $\sin^2(2\theta) = 0.75$.

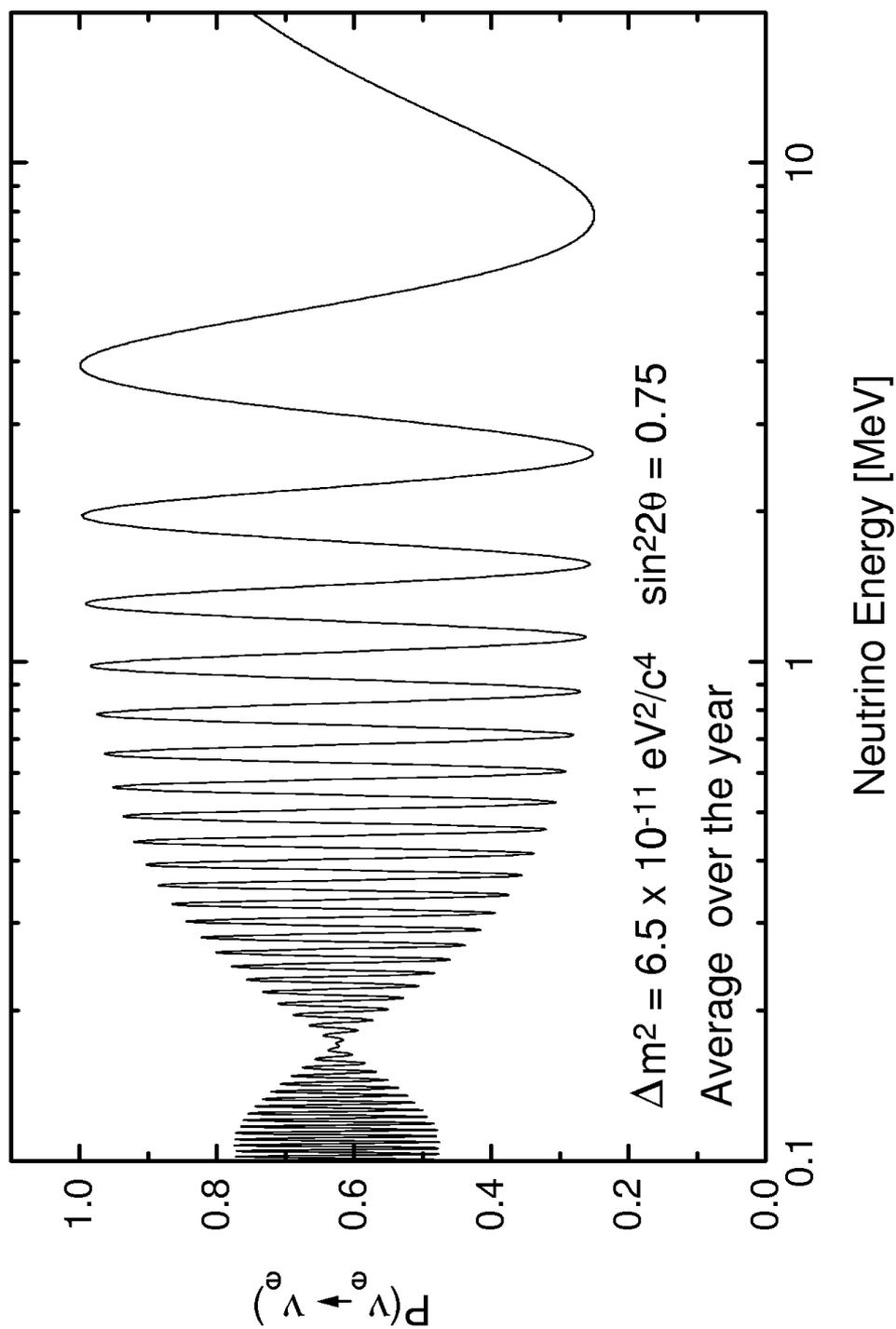

**Figure 4**: The $\nu_e$ survival probability due to neutrino flavor conversion, calculated for the three different MSW solutions: SMA (Small Mixing Angle), LMA (Large Mixing Angle), and LOW (Low probability, low mass). From Ref.[24] For updated explicit best fit parameters see caption of Fig. 5. At night time, flavor regeneration in the Earth can enhance the rate. Shown are yearly average survival probabilities:

total (solid lines); day-time (dotted lines); night-time (dashed lines).

The $^7$Be-$\nu$ line at 862 keV is almost completely suppressed in the SMA solution, as indicated by experimental data.

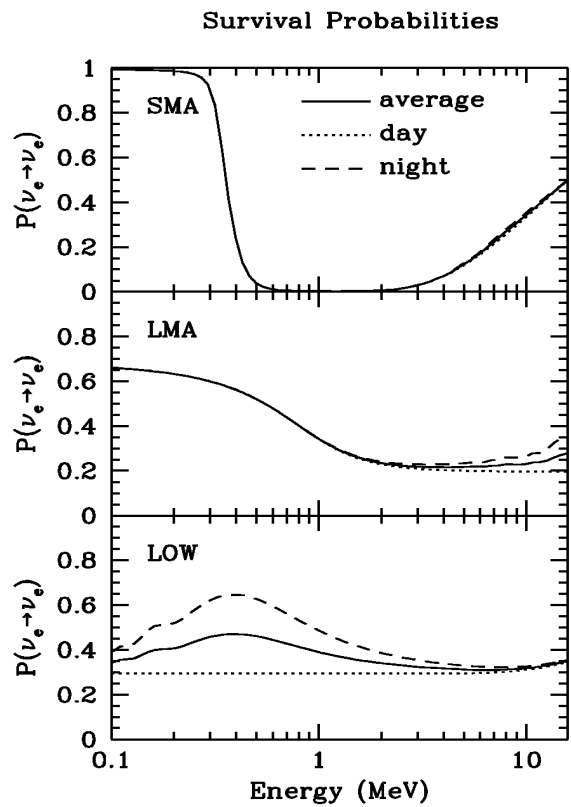

**Figure 5**: Plot of the allowed MSW solutions (99% C.L.) in the $\Delta m^2$-$\sin^2(2\theta)$ parameter space as deduced from the results of the Homestake, Superkamiokande, Gallex and Sage experiments. The best fit parameters are:

SMA (Small Mixing Angle), with $\Delta m^2 = 5.4 \times 10^{-6} (eV/c^2)^2$ and $\sin^2(2\theta) = 5.5 \times 10^{-3}$

LMA (Large Mixing Angle), with $\Delta m^2 = 1.8 \times 10^{-5} (eV/c^2)^2$ and $\sin^2(2\theta) = 0.76$

LOW (Low probability, low mass), with $\Delta m^2 = 7.9 \times 10^{-8} (eV/c^2)^2$ and $\sin^2(2\theta) = 0.96$.

Only the total event rates have been used in the calculations.[27] Bahcall and Pinsonneault 1998 = Ref.[5]

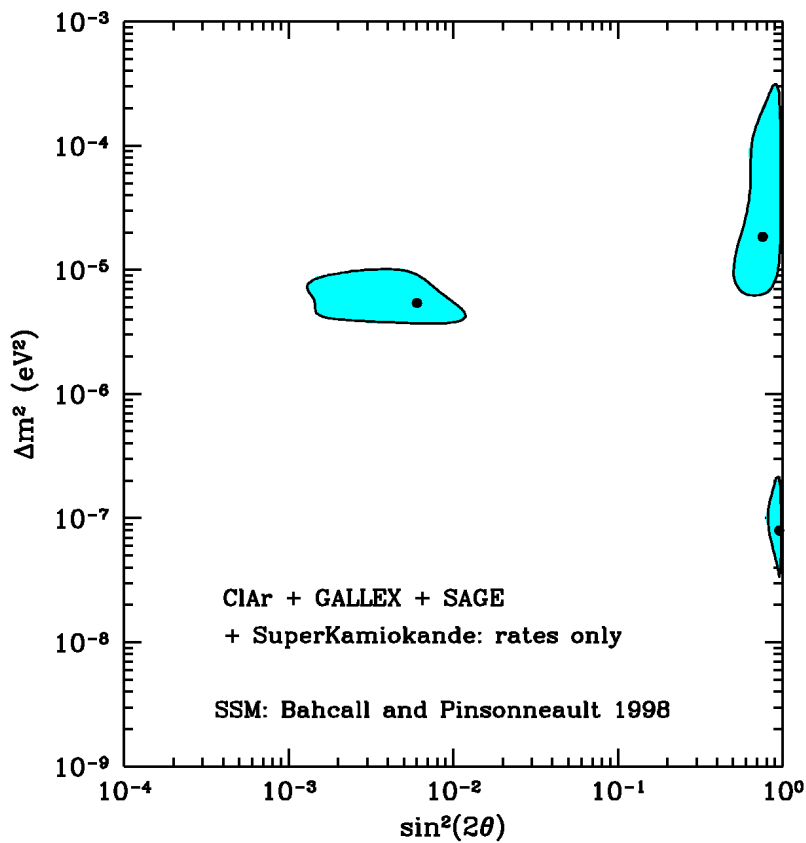

**Figure 6**: Plot of the allowed solutions in the $\Delta m^2$-$\sin^2(2\theta)$ parameter space for vacuum oscillations between active neutrinos as deduced from the results of the Homestake, Superkamiokande, Gallex and Sage experiments. Only the total event rates have been used in the calculations.[27] Bahcall and Pinsonneault 1998 = Ref.[5]

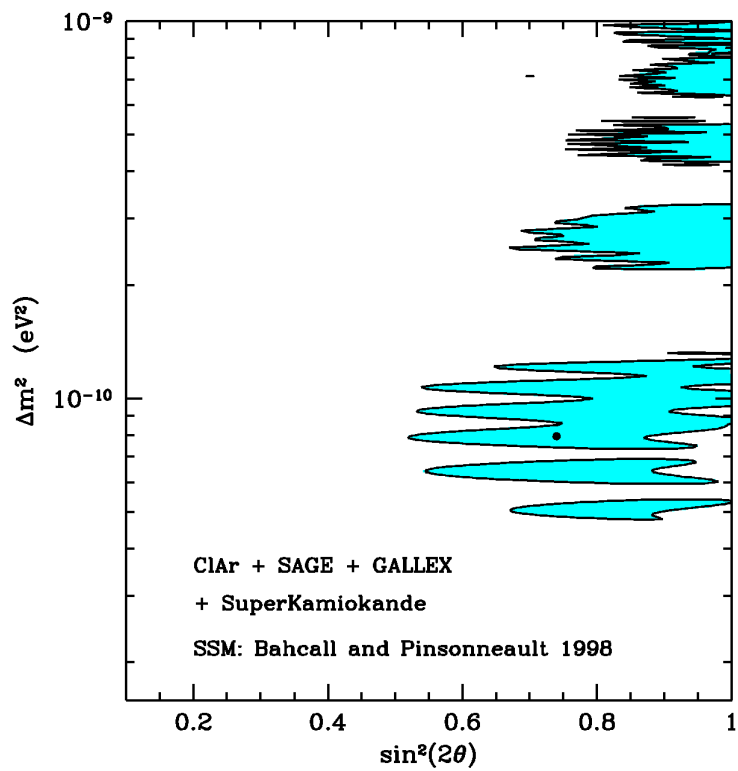

Neutrino-electron scattering is a purely weak process with a precisely known cross-section.[28] For the mono-energetic $^7$Be-$\nu$ line at 0.862 MeV, the profile of the recoil electrons is a unique 'flat box' with a spectral (Compton) edge at 0.66 MeV, determined by the $\nu$ line energy. The recoil profile and its edge energy are thus spectroscopic signatures for the $^7$Be-$\nu$ signal in Borexino. In principle, the shape of the profile is slightly different for $\nu_e$ and $\nu_{\mu,\tau}$. Unfortunately, the difference is probably too small to be useful in practice. Even though the recoil profile extends from zero to the edge, in practice the optimum $^7$Be-$\nu$ signal window in Borexino is (0.25-0.8) MeV, set mainly by background considerations ($^{14}$C). In this window, the nominal scattering cross-section is $\approx 5 \times 10^{-44}$cm$^2$ (at $\approx 1$ MeV). Thus, for a 100-ton detector and $\varphi_\nu(^7$Be) as in Table 1, an event rate of $\approx 18,000$/yr is expected, compared with a typical $\approx 100$/yr rate in the first generation experiments. Borexino is thus a *high rate* solar $\nu$ detector.

*Hard Core Signal Limit:* The $\nu_e$-e scattering is driven by the charged (CC) and neutral (NC) weak currents, whereas $\nu_\mu$ and $\nu_\tau$ scattering occurs only via the NC. The effective cross-section thus depends on the $\nu$ flavor. In the $^7$Be-$\nu$ signal window, the event rate due to the NC interaction alone is $\approx 20\%$ of the total rate (the NC/CC signal fraction at these low energies is $\approx 50\%$ higher than at $^8$B-$\nu$ energies). Thus, even for complete conversion of the $^7$Be-$\nu_e$ (likely according to the Ga results), Borexino will record a pure NC signal from just the converted $\nu_{\mu,\tau}$. The NC component thus sets a substantial 'hard core' *lower* limit on the solar signal in Borexino and effectively makes the experiment *sensitive to all $\nu$ flavors*.

*Sterile Neutrinos:* The discussion so far tacitly assumes conversion to normal 'active' $\nu$'s. In principle, conversion is possible to 'sterile' species i.e., neutrinos of the wrong helicity with interaction strengths reduced to vanishing levels by the factor $(m_\nu/E_\nu)^2$, so that they are effectively decoupled from weak processes. In this case, for complete conversion, the NC scattering is absent, thus the rate could drop even below the 'hard-core' limit of the $\nu$-e signal for active neutrinos, all the way to null levels. Such a result could be the signature for the presence of sterile $\nu$'s. In fact, this signature is unique to $^7$Be solar $\nu$'s because:

*(i)* A $^7$Be-$\nu$ signal below SSM-levels by a factor $\geq 4$ cannot be caused by *astrophysical* reasons, that being ruled out by the conflict in the $^8$B-$^7$Be signals.[27]
*(ii)* The global results of present experiments cannot accommodate a reduction by a factor $>3$ for any other feature in the spectrum except the $^7$Be-line.[27]

Thus a detector based on $\nu$-e$^-$ scattering specific to the $^7$Be flux, such as Borexino, has access to the window below the hard core signal limit for a *unique identification of sterile $\nu$'s* if the substantially more stringent background requirements can be met.

## 3. Technological Foundations

The electron recoil signal in Borexino from $^7$Be-$\nu$'s occurs in the energy window (0-0.66) MeV at a rate of 0.1-0.5 events per day and per ton of target material. Observation of such rare low energy events has never been attempted in real time because of the large un-shieldable background of the detector medium itself, arising from ubiquitous naturally radioactive contaminants such as $^{238}$U and $^{232}$Th. Radiations from these (and other) activities produce events at vastly higher rates in the entire range up to 5 MeV even in so-called high purity materials certified to ppt (i.e. parts-per-trillion) U/Th content. The only real-time observations of solar ν's to-date have therefore been made in energy regimes above ≈6 MeV.

$^{238}$U generates a background rate of ≈$10^5$Hz/gU. With this rate, a signal/noise ≈1 implies a contamination limit of $10^{-16}$gU/g. At the start of planning for Borexino, such purities had never been demonstrated even in laboratory samples, let alone on the multi-100ton scale. No chemical insight was available to assess whether such purities are realizable even theoretically, no chemical process has been demonstrated to achieve purification to such levels, nor were any analytical methods known for determining such ultra-trace amounts. The highest purity yet observed, in the case of water, fell short by orders of magnitude. While it was suggested that liquids less polar than water might be further purifiable, no measured data either precluded or promised the above purity levels.

Background from external sources is also severe, though removable in principle by massive shielding of the detection volume. In this respect, however, good precedents were available from water Čerenkov detectors such as Kamiokande. They showed that the external background can be suppressed by defining a fiducial detection volume in the core of a massive tank of pure water large enough to provide a shield thickness of ≈5m. To adopt this general approach to low energy spectroscopy, signal luminosity higher than that of the Čerenkov effect is required. A natural choice is the well known technique of liquid scintillation that offers ≈50 times larger signal luminosity and allows event sensitivity down to ≈50 keV. Such a strategy still involves two open problems:

*(i)* Containment of the LS in the shield liquid (buffer) sets severe constraints.

*(ii)* Although widely used in many areas in physics and biosciences, liquid scintillation had never been applied on the multi-100ton scale for spectroscopy at ≈50 keV.

Thus, a new technology had to be developed for observing low energy solar ν's in real-time.

### 3.1 Radiopurity

#### 3.1.1 Chemistry

The sources of radioactivity in the Borexino LS stem from:

*(i)* widely prevalent *primordial radioactivities:* $^{40}$K, $^{238}$U and $^{232}$Th and their decay products $^{226}$Ra and $^{210}$Pb;

*(ii)* ambient *noble gas activities:* $^{222}$Rn (radon) and $^{85}$Kr; and

*(iii)* *cosmogenic activities:* $^{14}$C, $^{7}$Be and $^{3}$H (tritium).

$^{14}$C is a special case that challenges the central concept of Borexino because its background critically interferes with the signal and it is *inseparable* from the organic liquid scintillator. All the other impurities are chemically separable, the question being the degree of removal. They can be grouped chemically as metals (U, Th, K, Ra, Pb, and Be) and noble gases (Rn and Kr). Tritium is not critical because of its low decay energy.

*Radiocarbon:* The β-spectrum of $^{14}$C ($E_{max}$=0.156 MeV) overlaps the signal window of pp-ν's (0-0.25) MeV and, because of the limited energy resolution in liquid scintillators, seriously tails into the signal profile of $^{7}$Be-ν's (0-0.66) MeV. At the start of Borexino, the most sensitive limit of $^{14}$C in any material was

$^{14}C/^{12}C < 10^{-15}$ which produces a $^{14}C$ rate $\approx 2 \times 10^7$/d,ton in an organic liquid, ruling out detectability of the much weaker signals from pp-v's ($\approx 2$/d,ton) and from $^7Be$-v's ($\approx 0.5$/d,ton). The latter can still be usefully measured in a window above 0.25 MeV if $^{14}C/^{12}C$ is at least as low as $\approx 10^{-18}$ ($\approx 10^{-6}$ of the $^{14}C$ in modern carbon). In the Borexino framework, this leaves no option but an organic material naturally 'free' of $^{14}C$. The only means to approach this goal is the strict use of *petrochemical* organics. The underground residence of such material for millions of years removes the original $^{14}C$, leaving, at worst, only small amounts of fresh $^{14}C$ from possible neutron reactions underground. These premises were encouraged by tests on isotopically enriched natural gas,[29] selected as a petrochemical proxy. Measured ratios $^{14}C/^{12}C$ $< 10^{-18}$ set the stage for tests of the actual scintillator petrochemicals in the CTF (see Sec. 3.2.3).

*Metallic Radioimpurities:* The purity regimes of Borexino are unprecedented (e.g., [K]$\approx 10^{-14}$g/g, [U,Th]$\approx 10^{-16}$g/g, [Ra]$\approx 10^{-22}$g/g, [$^7Be$]$\approx 10^{-26}$g/g) and inaccessible to the most sensitive analytical methods that offer at best, ppt sensitivities in favorable cases. For this reason we ultimately constructed a Counting Test Facility to explore low level backgrounds, particularly for U and Th (see Sec. 3.2). Before the CTF was constructed, however, a number of basic studies were made that provided insight on scintillator materials and some of their background problems. The studies also suggested methods for removing observed radioactive impurities, methods ultimately employed in the CTF. We mention briefly these studies on the scintillator solvent (initially TMB, later PC) and the wavelength shifter (PPO).

*(i)* Solvent studies (TMB and PC): Direct measurements of U and Th in the scintillator solvents could achieve levels of $\approx 1$ ppt by mass spectrometry. Efforts at concentrating impurities by analysis of residues after distillation led to limits of $\approx 10^{-15}$g/g for U and Th in tri-methyl borate, an early possibility for the scintillator solvent. While these studies did not reach the required sensitivities they did show that such solvents are quite pure as received, hinting at the possibility of the higher purities required for Borexino.

Radioactive tracer studies were also performed in which one tracks impurities rather than measuring their absolute content.[17,18,30] Instead of targeting impurity *content* with inadequate analytical sensitivities and uncertain blank values, one tracks impurity *transfer* in processes of purification, handling and storage. Starting from liquid scintillator materials containing U, Th, Pb, Ra and K at high levels typical of terrestrial abundances ($\approx 0.1$ to 1ppm) and spiked with tracers for these elements, ordinary processes such as distillation removed the impurities in the distillate by about 10 orders of magnitude. Useful efficiencies were obtained for water extraction. In addition, tracer work led to the discovery of a new purification technique. Laboratory tests showed that silica or alumina columns, normally used only for removing optical impurities, are also powerful devices for radio-purification. They removed impurities that had been initially in the liquid, and did not contaminate the solvents in so doing. In the case of $^7Be$ with the most stringent specifications (see above), pseudocumene (PC, chosen for Borexino) was activated directly by protons and neutrons (as in cosmogenic activation) and distilled, removing the $^7Be$ activity by at least three orders of magnitude.[30,31] Water extraction also produced useful Be purification.[31]

*(ii)* Wavelength shifter studies on PPO:[2,5-$(C_6H_5)_2(C_3HNO)$ = 2,5diphenyloxazole]: Commercial PPO was found to contain ppm concentrations of K that were readily detected by neutron activation analysis. Since the PPO is mixed with PC at concentrations of $\approx 10^{-3}$, the K impurity in the PPO would be at a level of $\approx 10^{-9}$ in the scintillator, or about 5 orders of magnitude higher than the requirement of $10^{-14}$g/g. Purification studies showed that the K could be readily removed from PPO by a water extraction process. Namely, the PPO is dissolved in PC at high concentration and the mixture is then vigorously contacted with high purity water. During mixing, the K is transferred to the water, because of its higher affinity for more polar solutions. After mixing, the water and the PC-PPO liquids phase separate (they are immiscible), and the K stays with the water, leaving the PC-PPO solution purified. Repeated application of the procedure reduces the K content in the PPO to an undetectable level. However, the detection sensitivity limit for K in the scintillator has not reached levels below $10^{-12}$g/g, neither by neutron activation measurement of PPO nor by direct spectroscopic measurements in the CTF. The purity requirement for K is thus not directly confirmed, but it is expected to be met because of the high efficiency of water extraction for removal of K. These studies led to the development of a water extraction column in the purification systems for the CTF and for Borexino.

The principal conclusion reached in the work above with several candidate scintillator liquids is that metallic impurities have a vanishing solubility in *non-polar* organic liquids. The extreme purities can be maintained indefinitely by removal of metallic impurities by ordinary chemical processes such as distillation and water- or solid column extraction. Liquid scintillators, in general, thus offer a practical route to low energy neutrino spectroscopy.

*Radon and Krypton:* Radioactive noble gases are mobile and chemically inert. They can be segregated using their volatility but not by chemical reactions. $^{222}$Rn (3.8d) emanates from Ra-decay in the $^{238}$U chain. The decay of $^{222}$Rn in the scintillator creates most of the radiation of the U chain and in addition, leaves a deposit of its end product, 22yr $^{210}$Pb$\rightarrow$$^{210}$Bi that produces background just in the $^7$Be-v signal window. $^{85}$Kr (10.85yr) is prevalent in the atmosphere and also produces background in the signal window. Thus, minimizing air leaks and Rn emanation sources in all parts of the detector are clearly major engineering issues with several facets (see below).

### 3.1.2 Analytics

*Active Spectroscopic Tags*: Solar neutrinos are detected in Borexino through their scattered electron events, without any distinct signature. However, a real-time identification is possible for several key types of background events that are caused by radioimpurities in the scintillator.[17] The tags are highly specific tools for background spectroscopy and play a crucial role in:
*(i)* the active suppression of most of the intrinsic background in the signal window; and
*(ii)* in the on-line quantification of impurities.

Untagged parts of the background from decay chain segments associated with the detected impurity can be reconstructed and statistically cut from the observed spectrum to the extent that they are in secular equilibrium with the directly identified species. The tags/statistical cuts can suppress typically $\approx 90\%$ of the background from U/Th and Kr-decays. The tags are of two kinds: pulse shape discrimination (PSD) of α-particles and delayed coincidences (DC) of correlated decays of impurity activities.

*PSD:* For every decay of $^{238}$U or $^{232}$Th, there is a sequence of 8 or 6 α-particles emitted, respectively, of energy ≈(4-9)MeV. They produce scintillation signals quenched by factors of 10 to 15 relative to β- or γ-signals, so that they appear in the 0.25-0.8 MeV energy window. Thus, a major background against the $^7$Be-ν induced e$^-$-signal is from U/Th α-particles. However, the signal quenching mechanism also results in characteristically long scintillation times that in turn lead to pulse shapes that differ from the much faster e$^-$/γ pulses. PSD can tag α-events with 90-99% efficiency, depending on the composition of the scintillator and on the α-energy.[32]

*Delayed Coincidences (DC):* Key components in the decay chains of U/Th and in the β-decay of $^{85}$Kr are emitted as time-correlated coincidence pairs which can be tagged with high specificity.[17] The β-α coincidences in $^{212}$Bi and $^{214}$Bi and the β-γ pair in $^{85}$Kr quantify the presence of the long-lived parent impurities. The $^{212}$Bi tag assays the entire segment following $^{228}$Th in the $^{232}$Th chain while $^{214}$Bi tags the $^{226}$Ra($^{222}$Rn)-$^{214}$Po segment of the $^{238}$U chain. The Kr and $^{214}$Bi tags are particularly useful in tracing the origins of persistent background from the gases, in particular, atmospheric leaks of Kr and Rn and internal source materials that emanate Rn.

*Trace $^{238}$U determination:* The $^{238}$U-$^{234}$Th-$^{234}$Pa-decays in the initial segment in the U chain do not offer DC tags or conveniently detectable α-particles. Equilibrium of $^{238}$U with the rest of its chain cannot be assumed *a priori*, since purification efficiencies are chemically specific. Thus, the specific determination of U in the scintillator is vital, at least sample-wise by off-line methods. The standard methods ($10^{-10}$-$10^{-12}$g/g sensitivities) fall far short of Borexino levels. A major innovation was made to extend the sensitivity of U-neutron activation analysis (NAA) by >6 orders of magnitude by a new technique, ISAN (Isomer Spectroscopy of Activated Nuclei).[33] The method is based on detecting impurities by activating them by a mode that creates nuclear isomers at least in one of the isotopes of the impurity. Isomeric decays can be specifically detected by DC tags. ISAN can be realized in practice, particularly for U: the reaction $^{238}$U+n → $^{239}$U → $^{239}$Np → (βγ DC) occurs naturally in normal neutron activation of U, needing only technical modifications of NAA counting for trace determination of U.[17] The technique is now developed for standard analysis of scintillation solvents with a trace sensitivity of ≈$10^{-16}$ to $10^{-17}$g/g.[33,34]

### 3.1.3 Optical properties

While not directly related to radioactive impurities, one should note the importance of non-radioactive impurities that can affect the optical clarity of the solvent. Such impurities, which are organic in nature, reduce the attenuation length of light passing through the scintillator, contrary to the requirements for a large liquid scintillator as in Borexino.[35] Degradation of the solvent can occur with exposure to air and exposure to certain metal surfaces. Distillation of the solvent was found to remove the impurities and restore the optical quality of the scintillator.[36]

*3.2 Counting Test Facility*

*3.2.1 CTF, a prototype for Borexino*

A 'Counting Test Facility' (CTF) was constructed and installed in Hall C of the Gran Sasso Laboratory. The goals of CTF were:

*(i)* to achieve ultra-purity in the scintillator on a massive scale;

*(ii)* to actually measure the prevalent radiopurity;

*(iii)* to perform spectroscopy at very low energies in a multi-ton detector;

*(iv)* to develop technical solutions for a host of engineering issues involved in achieving the above.

CTF is a large-scale test detector that is nonetheless modest in size relative to Borexino. A mass in the 4 ton range was set by the need to make the prevailing scintillator radioimpurities measurable via DC tagged events, while a water shield thickness of ≈4.5m was needed to not disturb this measurement by external radiation. The primary goal of CTF was to develop solutions directly applicable to operational issues for Borexino; but at the same time there was the long-range goal of performing quality control during Borexino operations.

Detailed reports on the CTF have been published.[37,38,39,40] In this section we outline only the main features and results. As a simplified scaled version of the Borexino detector, a ≈4m$^3$ volume of liquid scintillator is contained in a 2m diameter transparent nylon vessel (inner vessel, IV) mounted at the center of an open structure that supports 100 phototubes (PMT)[41] which detect the scintillation signals. The PMT's are coupled to optical concentrators viewing the IV with ≈20% optical coverage. The whole system is placed within a cylindrical tank (11m×10m) that contains ≈1000 tons of ultra-pure water which provides a shielding of ≈4m against neutrons originating from the rock and against external γ-rays from the PMT's and other detector materials. The CTF also includes systems for water and scintillator purification that consist of units for water extraction, vacuum distillation, silicagel column treatment, and for nitrogen sparging. The entire detector is built on clean-room construction standards.

A principal part of the CTF program is the development of technical solutions for achieving large scale LS ultra-purity. All parts of the detector contacting the LS - including storage, handling, purification and detector filling - met exacting standards to counter two basic modes of contamination:

*(i)* gaseous radioactivity (emanated Rn from Ra in materials and leakage of Rn, Kr and $^{14}CO_2$ from outside); and

*(ii)* dust and particulates (with specific activities $10^{10}$ times higher than Borexino tolerances).

The former requires that the entire detector and ancillary systems be engineered to high gas-tightness to rigidly exclude external air. The Rn background is itself relatively short lived, but deposited Pb-Bi daughter activity on the surfaces is long-lived. Because of the latter, Rn control is necessary not only in the detector and its operations but during materials production and assembly above ground. A rigid quality control program using underground Ge spectrometers and radon emanation tests screened all construction materials for radiopurity. The PMT glass was a special type with radioactivity ≈10 times lower than normal. The surfaces of the detector subsystems and of the plants, especially those contacting the LS were smoothed (to 0.6-0.8μm) and pickled/passivated or electro polished for effective cleaning of particulates.

*3.2.2 Major CTF Systems*

*The Liquid Scintillator (LS):* The scintillator used for the major part of tests in the CTF is pseudocumene (PC), widely used because of many desirable properties, e.g. high specific scintillation output ($\approx$12000 photons/MeV), long light transmission lengths (typically $\approx$7m) and commercial availability in kiloton quantities. In order to shift the emission wavelength to $\approx$380nm, to achieve long transmission lengths and to better match the peak of the photosensitivity of the PMT's, a fluor has to be added to the scintillator.[35,40,42]

The key issue to achieving both large transmission length and high photoelectron yield is auto absorption and re-emission of the flour. A research program on flours resulted in the choice of PPO at a concentration of 1.5 g/L scintillator. Some 5 tons of the LS were transported underground within 24-48 hours of production at the petrochemical refinery, thus minimizing cosmogenic production of $^7$Be during transportation at sea level. The PC was produced in the Enichem plant at Sarroch (Sardinia).

Tests were also carried out with an alternate LS solvent, phenylxylylethane (PXE), with comparable scintillation and optical properties but a specific gravity of 0.992, providing near-neutral buoyancy with a water shield. The flash point is $\approx$150°C, compared to that of PC at $\approx$45°C.

*Scintillator Containment Vessel (IV):* The design of CTF and Borexino calls for a highly transparent containment vessel for the LS in the core of a massive volume of shield liquid. Though this is simple in concept, in practice there are severe constraints on the design and materials. The material must be chosen for its strength, optical properties, radiopurity and chemical compatibility with the LS and shield liquids ('buffer'). The mechanical design must insure vessel integrity and allow feasible and cost effective construction. The key design innovation here is the choice of a thin membrane containment vessel, rather than a rigid shell.[43] As one of the few choices available for maximum compatibility with the various constraints, thin (0.5mm) nylon film was selected as the vessel material. In addition to the need for high bulk radiopurity, one must also face the reality that dust and aerosol-borne Rn daughters can be deposited on the nylon surface during film manufacture or balloon fabrication with the possibility of contamination of the LS over time.[43] Extrusion of the nylon as well as construction of the IV were carried out under clean room conditions, with special efforts to reduce the plateout of radon daughters. The density of PC is $\approx$0.88g/cm$^3$, well below that of water. Nevertheless, the concept of membrane containment of the PC LS in water was found to be mechanically sound at the 4-ton level and the CTF vessel was operated safely for two years.

*Water Purification:* This system consists of a filling plant and a re-circulation loop.[44] Pre-filtered (0.1μm) water passes through a reverse osmosis unit, a continuous de-ionization unit and finally is sparged by counter current flow of nitrogen in a stripping column.

*Scintillator Purification:* The CTF scintillator purification system is designed to perform pre-purification of the scintillator and its components and on-line purification after the scintillator is installed in the detector. Particularly important is the removal the impurities that are known to be present in the fluor and in the quencher materials. We have two major plants for both, 'off-line' (pre- or batch-) purification and for 'on-line' purification[36,45]:

*(i)* The PPO-fluor and DMP-quencher are pre-purified in concentrated PC solutions using the solid column and batch purification plant. It allows batch operations for water extraction, nitrogen degassing and sparging, as well as ultra-purification and filtration by means of chemical packings. The plant has been generally tested during CTF operations and used to prepare the scintillator mixtures thus-far reported. It can also handle on-line purification of post-loaded scintillator solutions as well as PC (fluor free liquid).

*(ii)* The second plant consists of sub-micron filtration, water extraction (of ionic species), vacuum distillation in a six plate system and nitrogen stripping (of noble gases such as Rn and Kr, water, and active gases like $O_2$ and $CO_2$). This plant handles purification of PC solutions in both on and off-line modes. The distillation subsystem is used on PC (fluor-free) liquid and has the additional feature to remove the fluor from the scintillator. The water extraction feature allows for continuous on-line water extraction of the post-loaded scintillator solutions.

### 3.2.3 Results

The CTF has set new milestones in at least two respects:

*(i)* It is the largest nuclear detector ever built with sensitivity down to the sub-100 keV spectral regime; and

*(ii)* It has the lowest background at 0.03 events/kg,keV,yr in the (0.25-2.5)MeV energy window, even lower in rate and in the energy regime as the presently best detectors for double beta decay.

Detailed results from the CTF are summarized elsewhere.[37,40] Here we outline only the results relevant to the new technology for Borexino.

*Signal Spectroscopy:* A signal yield of $\approx$300 photoelectrons/MeV was measured in the CTF scintillation detection system, permitting for the first time, nuclear spectroscopy down to a threshold of <50 keV in a multi-ton nuclear detector. Fig. 7 shows the β-spectrum of $^{14}$C ($E_{max}$= 156 keV) in the region (20-350) keV. Indeed, the spectrum is precise enough to test the weak-interaction shape of the spectrum.[38] The relative timing of the PMT hits allowed an event position resolution of $\approx$12cm (1σ) for the $^{214}$Po α-line (at the quenched energy of 750 keV). A PSD efficiency of $\approx$97% was achieved for $^{214}$Po α-identification with a loss of $\approx$2.5% of β-events (Fig. 8). New effects in the transport of scintillation light in massive detectors due to overlapping emission/absorption bands were identified and their effect on pulse height and timing was studied.[40]

*Impurity Spectroscopy and Radiopurity:* Events as singles and as tagged by DC led to the identification of the lowest contamination levels of the principal impurities ever recorded in any material. From the $^{14}$C spectrum above, the rate leads to $^{14}$C/$^{12}$C = (1.94±0.09)×10$^{-18}$ [38], close to the design goal of Borexino and establishing quantitatively its basic design premise.

The $^{226}$Ra content in the LS was determined by the $^{214}$Bi-$^{214}$Po DC tags which measure the Rn segment of the U chain. The $^{238}$U content in the LS was determined by off-line ISAN-NAA measurements of the $^{238}$U itself. The lowest achieved rate for the former was $\approx$1.5 events/d in the CTF, corresponding to (3.5±1.3)×10$^{-16}$g/g of $^{238}$U-equivalent. This is an upper limit since part of the signal was due to Rn from the water that migrated through the nylon vessel into the LS.

**Figure 7:** $^{14}$C β-spectrum between 20 keV and 350 keV measured in the CTF. The endpoint energy of $^{14}$C is 156 keV.

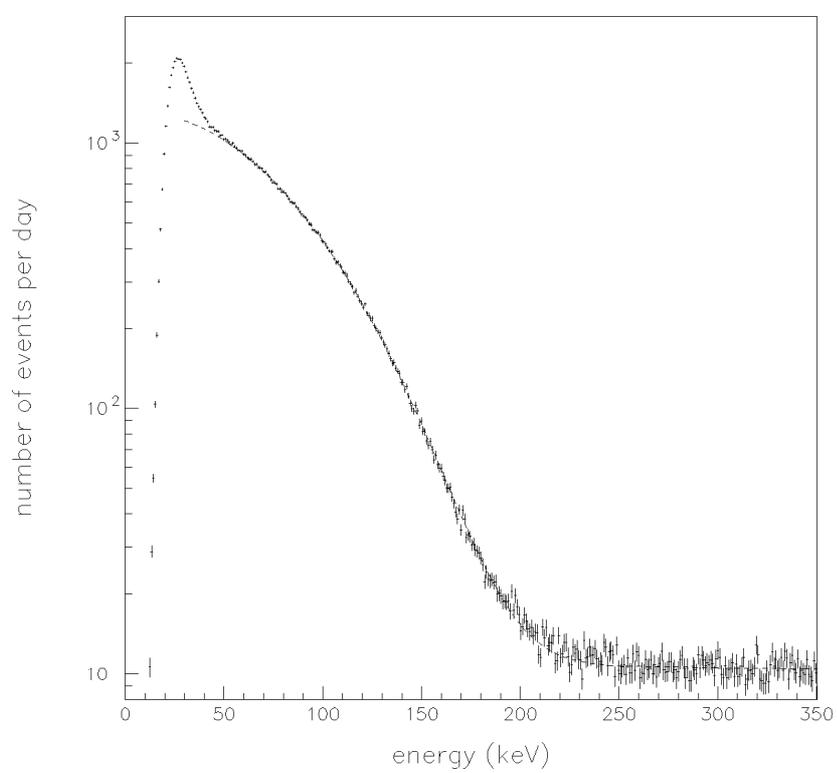

**Figure 8:** α-β separation in the CTF based on pulse shape differences demonstrated with coincident pulses from $^{214}$Bi (β-continuum up to 3 MeV) and from $^{214}$Po (α, quenched to 751 keV). The parameter plotted on the ordinate is (anti)correlated to the relative proportion of charge contained in the steep initial part of the signal (first 48 ns from a total of 500 ns). For detail, see.[37]

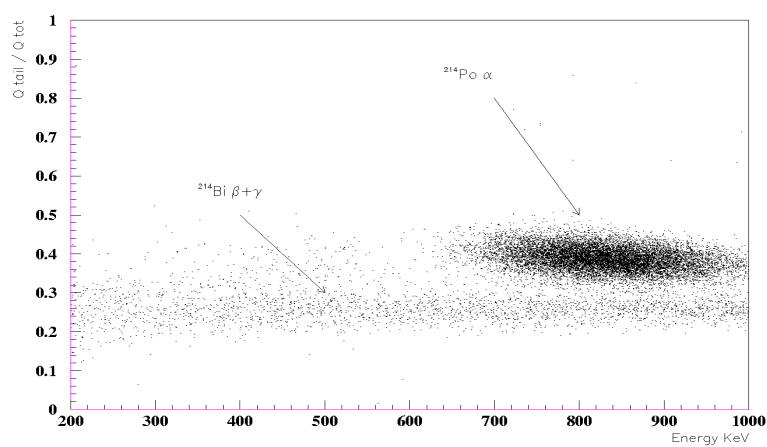

**Figure 9:** Energy distribution of the events tagged as $^{85}$Kr-candidates in the CTF. The distribution is compatible with the expected β-spectrum (endpoint energy 173 keV).

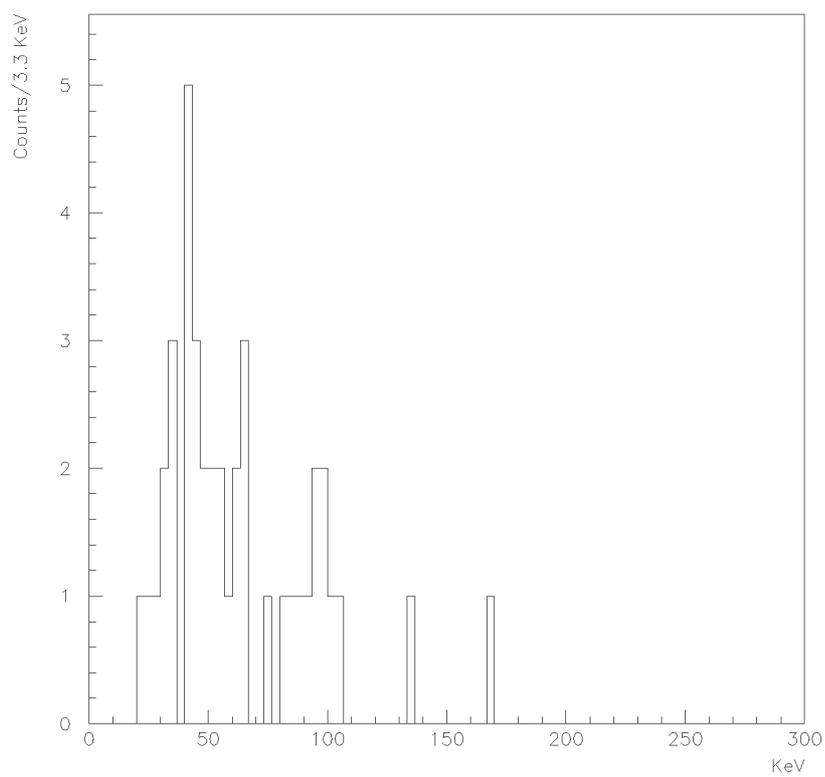

**Figure 10:** Detection of neutrons emitted by muon interactions. *Left:* Delay time distribution between the muon interaction and the neutron capture. *Right:* Energy spectrum of gammas after neutron capture on protons. Recognition of such events is required for the discrimination of the cosmic ray muon induced residual background.

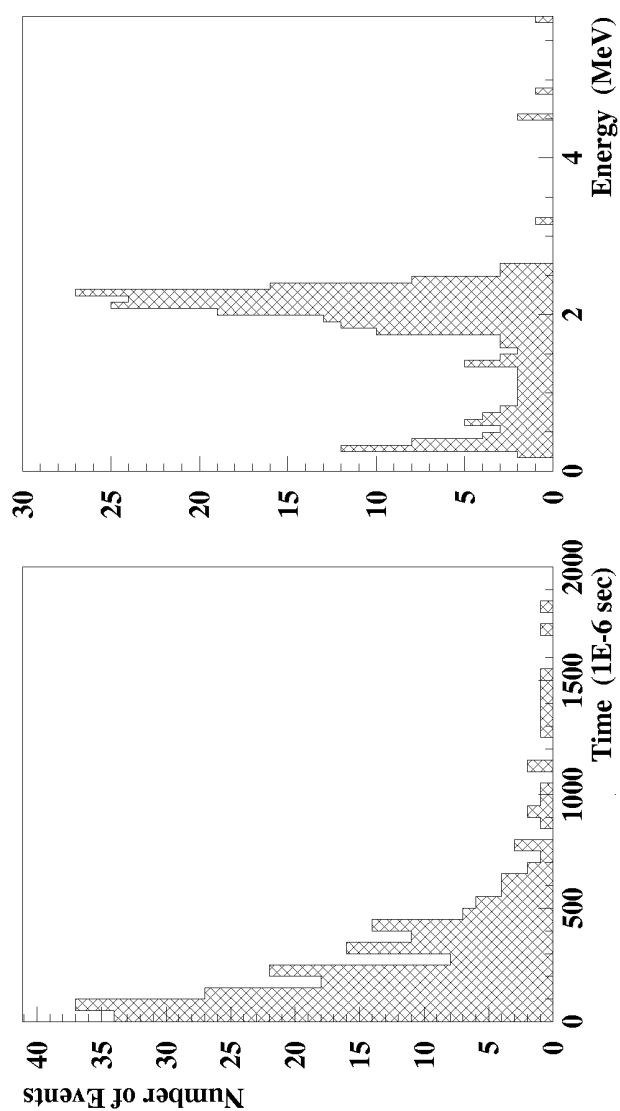

The Th-equivalent contamination was identified by the $^{212}$Bi-$^{212}$Po DC tags observed at the lowest rate of (0.31±0.10)counts/d corresponding to (4.4±1.4)×10$^{-16}$gTh/g. These values were observed in the PC *as received from the refinery* without any purification (with only the addition of ≈1.5g/L of purified scintillator fluors). ISAN-NAA analysis (performed at a later stage after operating the distillation cycle) gave <2×10$^{-16}$g$^{238}$U/g[34], limited by the method rather than by the sample's U-concentration.

*Radon and Krypton:* One of the most important contributions of the CTF to Borexino technology is the experience acquired on the radon problem, both in the scintillator itself as well as external to the scintillator, mainly in the water shield. Infusion of radon into the scintillator occurred episodically in large scale during various operations of liquid handling as well as at persistent low levels, possibly via permeation through the nylon vessel and during the sparging of the scintillator with large amounts of nitrogen. At the time of these experiments, the methods used to produce Rn-free nitrogen were not yet sufficiently effective. Meanwhile this has been drastically improved to a record purity level of ≈0.5μBq/m$^3$ (see Sec. 4.2). The leakage of Rn in the atmospheric air could be distinguished by detection of the accompanying $^{85}$Kr via its DC tag. Fig. 9 shows the tagged spectrum of leakage traces of $^{85}$Kr.[46]

*On-Line Purification of the LS:* All the installed on-line systems were operated successfully but revealed the inherent possibility of radon infusion, stressing the necessity of Rn removal by nitrogen sparging following the purification process. These operations also showed the removability of accumulated Rn by-products - $^{210}$Po and the daughter activities of Bi and Pb. Several cycles of water extraction removed these impurities satisfactorily (by at least ×20).[36,38] The performance of the solid column purification was evaluated in the CTF campaign with PXE (doped with ter-phenyl and bis-MSB fluors) by (now improved) off-line ISAN-NAA techniques. The results showed the achievability of ultimate ultra-purity in an aromatic solvent with the results:[$^{238}$U]<1×10$^{-17}$g/g and [$^{232}$Th]<2×10$^{-16}$g/g, compared to [U]=(2.7±1.5)×10$^{-14}$g/g and [Th]=(3.2±1.6)×10$^{-14}$g/g initially observed in PXE.[34]

*Purity of Shield Water:* The water produced by the CTF plant achieved the design goals with the following impurity results as measured by ICP mass spectrometry (U/Th: Ispra), by emanation techniques (Ra, Rn: MPIK Heidelberg) and by chromatography (K: Bell Labs): [U,Th]≈10$^{-14}$g/g; [K]≈10$^{-12}$g/g. The U,Th and Rn values are 4 and 6 orders of magnitude respectively, lower than for the raw Gran Sasso water. Measurements of Ra (long-lived parent of Rn) gave <20mBq/ton initially and <2mBq/ton after a recent upgrading of the assay system.[44,47,48]

*Cosmic Ray Interactions:* Interfering background due to passage of cosmic ray muons in the cave was studied mainly by two methods: identification via the PMT 'hit pattern' of the event, and pulse shape analysis. The data showed that as much as ≈2% of such events could be mistaken for low energy events in the scintillator.[37] More information was obtained from an external muon veto detector with a relatively small coverage which sampled the effect of muons in producing events at low energies correlated with neutrons produced by spallation. Fig. 10 shows the characteristic 2.2 MeV γ-rays from proton capture of diffusing neutrons in correlation with the muon signal. These measurements directly determined the rate of cosmic ray spallation background on-line at low energies for the first time as 0.3/d,ton of material (see Sec. 4.4.3). The results stressed the need for identifying muon signals with an efficiency of ≈99% and provided the main thrust for designing an efficient muon veto for Borexino.

*CTF-II:* Several questions raised during operations in the CTF are being revisited in an upgraded CTF-II facility designed to substantially reduce the operating singles rates encountered in CTF-I, due mainly to a high radon concentration in the shield water. The chief upgrades are a nylon screen between the scintillator vessel and the PMT's and a larger (but still partial) muon veto by upward looking PMT's arranged over the entire bottom floor of the detector. CTF-II is planned as the main quality control facility for the LS and for the operating systems of Borexino.

## 4. The Borexino Detector

### *4.1 Detector Systems*

Borexino is being constructed in Hall C of the Gran Sasso underground laboratory (LNGS) in Italy (Fig. 11) with an overburden of ≈3500 meter water equivalent (m.w.e.). The suppression of the the cosmic muon flux is roughly six orders of magnitude, down to a value of $1.1/m^2$,h.

The architecture of the detector (Fig. 12) basically aims at reducing the γ-ray background at the central fiducial volume (FV) to a level well below the minimum solar ν signal via graded shielding by various thickness of increasingly radio-pure materials as the FV is approached. The final layer of shielding is the outer part of the scintillation liquid itself (the active buffer) contained in a membrane balloon inner vessel (IV). Software selection of events is restricted to the FV, creating in effect, a 'wall-less' counting device with background limited mostly by the scintillator radioactivity. The IV is immersed in a high purity non-scintillating (inactive buffer) liquid contained by an outer vessel made of stainless steel (OV) which also supports the PMT array. The whole arrangement is immersed in turn, in high-purity water contained in the main tank. This graded shield architecture provides a total shield thickness of ≈5 m.w.e. which suppresses the external γ-ray and neutron backgrounds by some 10 to 11 orders of magnitude. The central part of the detector is the active mass of LS. The scintillation light is viewed by 2200 PMT's. Outward looking tubes on the surface of the steel sphere act as a muon veto detector using the Čerenkov light produced by muons crossing the outer water buffer.

*Liquid Scintillator and Inner Vessel:* The choice for the LS in Borexino is PC with ≈0.17% PPO added as in the CTF. The scintillation pulses in PC are relatively fast with a decay time of ≈3.5ns for β-particles; however, the effective decay time is broadened to ≈5ns because of absorption/re-emission phenomena as the photons propagate over long path lengths in the detector.[40,42] The LS is capable of excellent PSD as tested in the CTF. The total mass of LS is ≈290 tons contained in an IV 8.5m in diameter. The software defined FV diameter is nominally 6m, containing ≈100 tons of LS acting as the solar ν target and as detector. A larger fiducial volume may be possible with the low external background expected in the present design. The IV, made of 0.1mm thick transparent nylon, is tethered by a system of strings.[49] The string tensions will be monitored. A second nylon enclosure placed near the PMT shell minimizes inflow of radon and other impurities that may diffuse into the vicinity of the IV from the outer parts of the detector and produce γ-ray background in the FV.

**Figure 11**: Layout of Hall C in the LNGS.

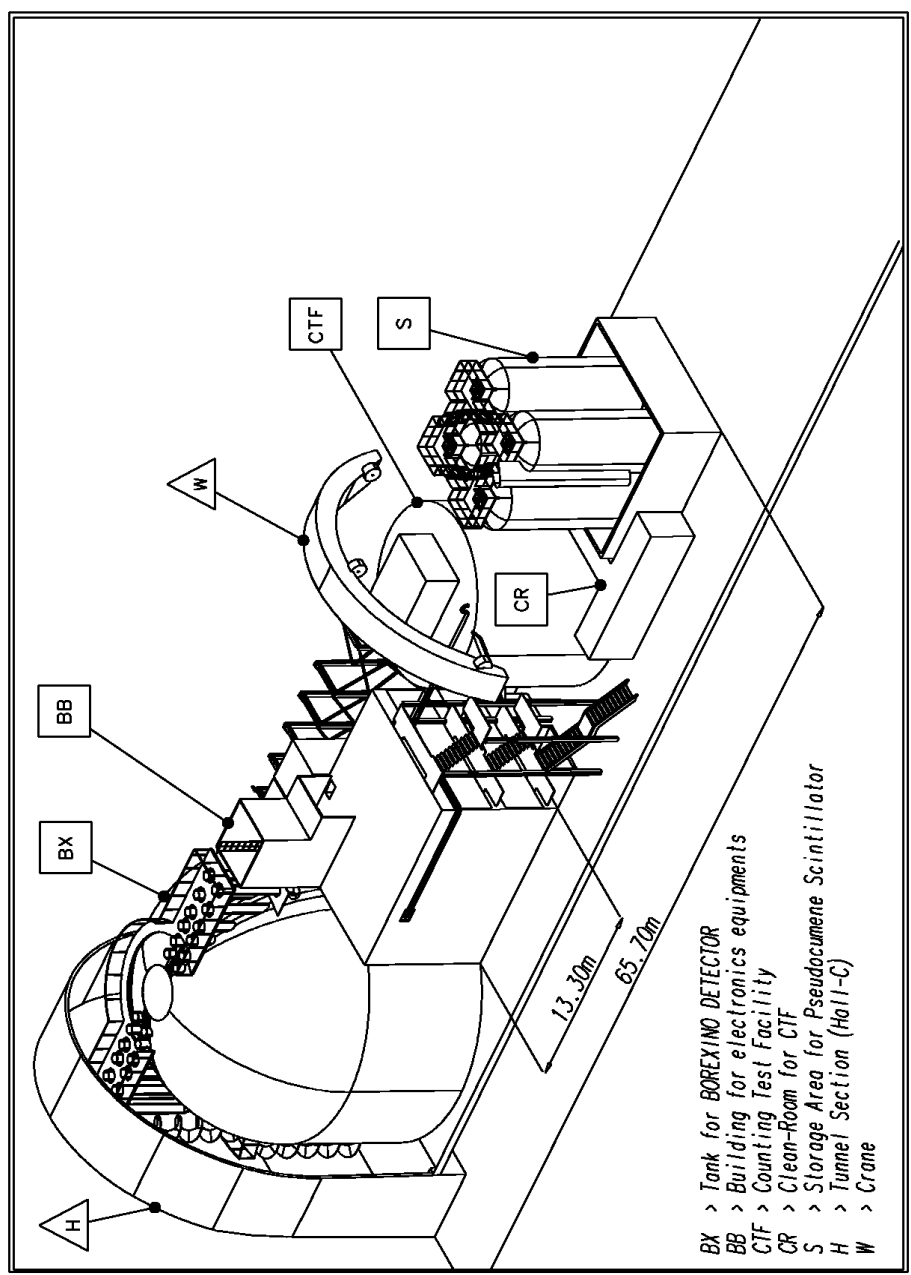

**Figure 12**: Sketch of the Borexino detector. About 300 tons of liquid scintillator are shielded by 1040 tons of a transparent buffer liquid. The scintillation light is viewed by 2200 PMT's. Reconstruction of the position of point-like events allows the determination of a 100 ton fiducial inner mass - the solar neutrino target. Outward looking tubes on the steel sphere surface act as muon veto detector. They use the Čerenkov light produced by muons that intersect the outer water buffer. The latter serves also as a shield against external radiation.

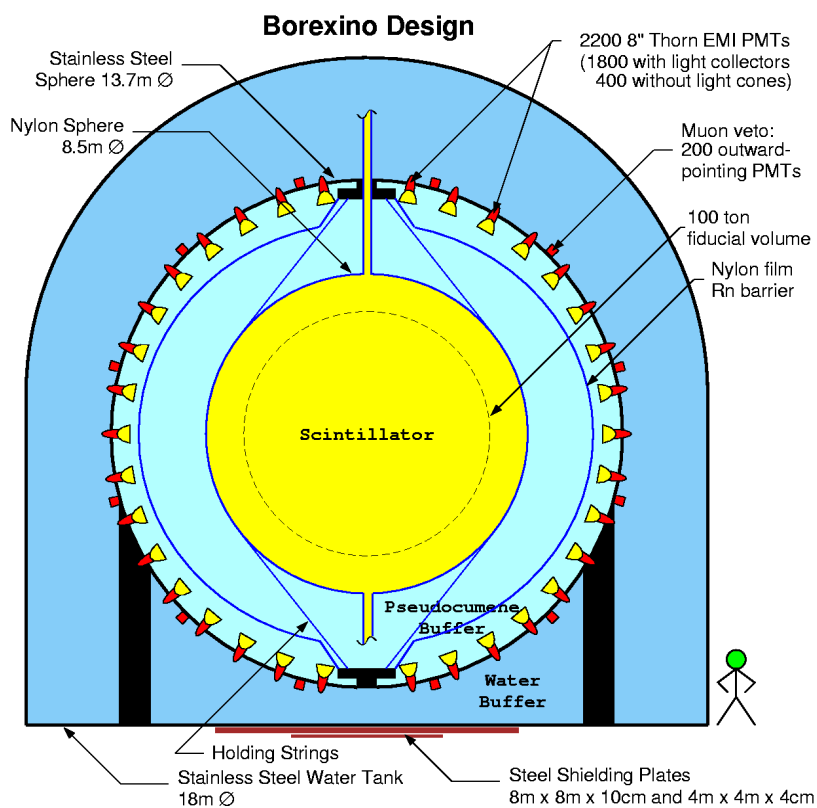

*Outer Vessel and Inactive Buffer:* The OV is a stainless steel sphere 13.7m in diameter and serves two purposes:

*(i)* it is the mechanical structure that supports the buoyancy force of the lighter scintillator and buffer compared to water, and

*(ii)* it is the support base for the PMT's.

The OV contains a transparent buffer liquid having a mass of 1040 tons and a density very close to that of the LS in the IV. The basic aim in the design is a *non*-scintillating buffer liquid that achieves a buoyancy condition as close to neutral as possible to minimize mechanical stresses on the IV balloon. The choice is PC but without the PPO fluors. However, even in the absence of fluors the PC itself produces scintillations in the ultraviolet. This is a potential problem since the PMT's, which are the most radioactive part of the detector, sit in the buffer. Because of the possibility of imperfect position reconstruction, some of these events may appear in the fiducial volume, producing a background in the ν signal window. Thus, a 'quenching' compound (dimethylphthalate, DMP)[50] is added to the buffer PC at a concentration of ≈5g/L to suppress the buffer fluorescence without affecting the transmission of the IV scintillation light. The DMP is also chemically compatible with the materials of the OV and PMT's.

An alternate buffer design that was considered is based on a water shield (as in the CTF) using PXE as the scintillator in the IV. The density of PXE is only ≈0.5% off that of water at typical operating temperatures. The density match could, if necessary, be improved by small amounts of suitable additives. The scintillation and optical properties of PXE are very similar to those of PC.

*Phototubes, Light Concentrators and Signal Quality:* The OV sphere supports 2200, 20cm ETL PMT's[41] installed inside the sphere, connected by feed-through's across the OV wall to a single cable outside the sphere carrying both signal and high-voltage. The back-end sealings of the PMT's have been designed to be compatible for operation in PC or water buffer. The PMT's have been selected for low radioactivity glass, low dark pulse rate, low after pulse rate, and a (1σ) transit time spread of ≈1ns, much smaller than the scintillation pulse-width. About 1800 of these tubes are equipped with light concentrators[51] to enhance the optical coverage, expected to be effectively ≈30%. Signals from the remaining 400 PMT's without light concentrators will be used for distinguishing muon tracks in the buffer and point like events in the scintillator (see below). The shape of the concentrators is designed to produce a uniform response to every event in the IV with a misalignment tolerance of ≈5 degrees. The PMT and concentrator system is rated to produce an average signal efficiency of 400 photo-electrons/MeV that translates to an energy resolution of ≈5% (1σ) at 660 keV. The resolution is suitable for observing the basic spectral signature of the Compton like edge of the recoil electron profile from [7]Be solar neutrinos. The system position resolution, determined mainly by the scintillation rise time and by the photoelectron yield, is expected to be <10cm (1σ) in each orthogonal direction at 1 MeV.

*Muon Veto:* Even though the underground location suppresses the cosmic muons, the CTF tests showed that muons that intersect the detector produce a prompt background of low energy events in the ν signal window.[39] These events arise mostly from muons that intersect the non-active buffer region and produce Čerenkov- and scintillation photons. Despite the presence of a quencher in the PC buffer, high levels of photons generated by the muons are not completely suppressed. To reduce this background to <1 event/d, an efficient muon veto system has been designed.[52] The Borexino muon veto consists of an 'outer detector' of 210 PMT's on the outside of the OV sphere and an 'internal detector' with ≈400 PMT's without concentrators. The former acts as water Čerenkov detector for intersecting muons. The latter accepts photons basically from all directions (unlike signal PMT's with concentrators that accept photons mainly from the active scintillator region). By comparing the total pulse height of both classes of PMT's, muon tracks in the non-active buffer region can be separated from point-like events in the active scintillator. In addition, the different time pattern created by muon events in the inner detector compared to point like β-signals can be used to separate background events. The overall design has been optimized in order to establish a redundant muon veto system to suppress this background by a factor of ≈$10^4$.

*External Tank:* Overall containment of the detector is provided by an external tank 18 meter in diameter and 17 meter high, filled with water as shield against external γ-rays and neutron radiation. In addition, the water acts as a Čerenkov medium for the muon veto system described above. An external plant (used in the CTF) feeds ultra-pure water to the tank. This ensures high optical transparency for Čerenkov operation.

### 4.2 Ancillary Plants

*Scintillation Fluid Management:* An elaborate system is under construction to store, handle and purify some 300 tons of PC scintillator and 1000 tons of PC buffer liquid. The main components of the system are: four storage tanks, four purification systems, detector filling systems, and the two detector systems of CTF and Borexino. In addition, there is an interconnection system for distributing the 300 tons of scintillator from the storage tanks into the Borexino detector and/or the CTF via or bypassing the purification systems. The system is made of electro polished stainless steel plumbing and high quality valves and fittings (He leak rate <$10^{-8}$mbar L/s), consistent with high purity chemical methods and the exclusion of radon and krypton.

On-line purification of the LS is based on four systems: gas removal, water extraction, distillation, and solid column chromatography to remove non-gaseous contaminants. Ultra-pure nitrogen is used to scrub Rn or Kr. Cosmogenic $^7$Be is produced during sea-level exposure to cosmic rays in the significantly long periods between large-scale production and storage underground. It creates a critical background orders of magnitude larger than the ν signal. One task of the on-site distillation facility (possibly also of the purification column) is to remove the cosmogenic $^7$Be contamination; the other is general purification for removal of optical impurities and other radioactive contaminants.

*Water Purification System:* Some 2000 tons of ultra-pure water must be produced and maintained to serve as the outer shield and as a highly transparent

Čerenkov medium. Another 1000 tons are required for operating the CTF. In addition, water is used for cleaning the detector and the scintillator fluid handling system and for ultra-purification processes such as water extraction for the scintillator. The system has a production capacity of 2 tons/h. The performance of the system is summarized in Table 2.

*Nitrogen Plant:* The nitrogen plant consists of three liquid nitrogen storage tanks, $6m^3$ each, two atmospheric evaporators and a water bath electric heater to produce $N_2$ gas of up to $250m^3$/h. Pure nitrogen gas is used in great quantities in handling the liquid systems, performing the final $N_2$ sparging step in the LS purification and in saturating the scintillator with nitrogen for oxygen-freedom (necessary to maintain scintillation efficiency). In order to reach the desired Rn purity in the *scintillator* (order of $1\mu Bq/m^3$), the nitrogen must contain less than $\approx 1\mu Bq$ $Rn^{222}/m^3$ $N_2$. At the time, this was a factor $\approx 100$ beyond state-of-the-art boil-off techniques. Technology was thus developed for large scale ultra-purification of $N_2$. High purity nitrogen ($1\mu Bq/m^3$ due to $^{222}Rn$) is produced by charcoal column purification of the liquid $N_2$ *prior* to evaporation. Up to 100 m$^3$/h of high purity $N_2$ can be supplied via electro polished lines equipped with high quality valves. Quality control of these gases is done after Rn concentration on charcoal with miniaturized radon proportional counters. The monitoring sensitivity is a record $\approx 0.5\mu Bq/m^3$ for $^{222}Rn$.[48,53]

### *4.3 Operational Elements*

#### 4.3.1 Signal Processing and Data Acquisition

The basic observables for the identification of $\nu$ events in Borexino are the total energy released in the scintillator, as measured by the number of photons emitted, and the time distribution of these photons. The electronic signal processing scheme shown in Fig. 13 is designed to achieve the timing properties needed for a variety of key tasks: reconstruction of the event position, PSD of $\alpha$- and $\beta$- types of events, and the identification of a variety of delayed coincidence tags with a wide range of time bases.

| $^{238}$U | $10^{-3}$ | $10^{-6}$ | $10^{-7}$ |
|---|---|---|---|
| Conta-mination | Raw water | Borexino design | Achieved |
| $^{226}$Ra | $3\cdot10^{-1}$ | $10^{-6}$ | $\approx10^{-6}$ |
| $^{232}$Th | $10^{-3}$ | $10^{-6}$ | $10^{-7}$ |
| $^{40}$K | $10^{-3}$ | $5\cdot10^{-6}$ | $<2\cdot10^{-6}$ |
| $^{222}$Rn | 10 | $10^{-6}$ | $\approx3\cdot10^{-6}$ |

**Table 2**: Summary of water radiopurity purification performance. Activities are given in Bq/kg.

**Figure 13**: Block diagram of the Borexino electronics layout.

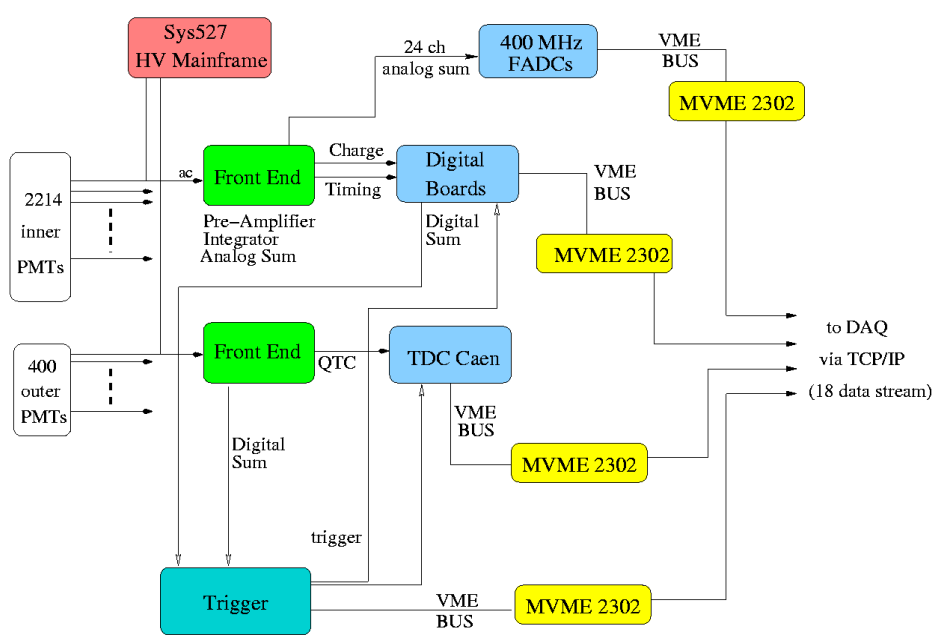

The data acquisition (DAQ) is based on the Linux operating system. The DAQ software is entirely custom made, with extensive use of multi-tasking techniques. User interfaces are all based on WEB techniques.

The signal from PMT's is AC coupled to a front-end card[54] that performs noise filtering, pre-amplification, shaping and integration of the input signal. It provides both a linear response used for time measurement and a voltage signal proportional to the total charge. Each front-end board provides also an analog sum of 12 linear output signals that extend the dynamic range of the system to ≈30 MeV by means of a flash ADC system. The outputs of the front end cards are sent to a specially designed VME slave card that performs the analog to digital conversion of the charge signal (8 bits resolution), measures the time of the linear signal with 0.4ns resolution, computes the sum of recorded hits in a time window of 60ns (used for triggering) and stores the whole information in a dual port random access memory.

The outer muon tubes are read with a different front end system that performs a charge to time conversion of each signal after a linear pre-amplification. The converted signal is then sent to time-digitizers which are read by their own processor. The trigger can be generated both by the internal as well as by the outer muon detector. We require at least $N$ PMT hits occurring in a time window of 60ns in the inner detector to generate a trigger. The trigger threshold $N$, determined by the operating singles rate, will ultimately depend on the $^{14}$C contamination, expected to be the largest contributor to the singles rate. In the inner detector, $N$ is expected to be 15-20, corresponding to (40-50) keV energy deposition.

### 4.3.2 Calibration and Monitoring

Even with an ultra-pure LS ($10^{-16}$g/g) that leads to very low background, determination of the detection efficiencies is vital. For example, the α-β separation efficiency remains critical especially within the ν window since α-particles of 4 MeV are 10 times more prevalent than the signal. With an anticipated separation efficiency of ≈90%, any energy or positional dependence on the separation would strongly impact the spectral interpretation and the inferred ν flux. In addition, the application of a number of software cuts, a basic one being the determination of the FV on a continuous basis, require the determination of precise efficiencies. An accurate calibration of the detection system is thus of paramount importance for quantitative precision, stability and reliability of the spectroscopy. The calibration program[55] covers the energy and the position sensitivity of the detector using active tags of trace impurities in the LS as well as external point sources inserted periodically into the tank.

*Laser Monitor:* The pulse timing and the gain of each individual phototube of the inner detector are calibrated by a laser system. Photons from this source are distributed to all PMT's via thin quartz fibers connected to the optical concentrators. The light yield corresponds to single photoelectrons as in real ν events. The outer muon veto detector will be calibrated by a set of blue light-emitting diodes that are mounted on the inside wall of the outer tank. Their gains match typical photon yields of Čerenkov events.

*Native Dispersed Radioactive Sources:* The energy response can be continuously monitored using the internal trace radioactivities native to the LS as well as selected inserted sources. The spectroscopic features of the sources together cover the entire range of energies up to ≈5 MeV that is of interest to most of the physics questions addressed by Borexino. A monitor for low energy short- and long-term stability is offered by the relatively high rate β-spectrum of $^{14}$C (0-156) keV. Even at a concentration of $10^{-17}$g/g U and Th, the β-α DC tags can be observed at a rate of ≈100 events/month and the mono-energetic α-particles at an even higher rate. These features can be utilized as long term monitors of the energy resolution and spectral stability while the spatial coincidences of the DC tags can be used to monitor the operating position resolution. Finally a calibration point at a higher energy is offered by the neutron capture γ-rays at 2.2 MeV which can be tagged in delayed coincidence triggered by the muon that produces the neutron. The CTF showed that these occur at a rate of 0.3 neutrons/ton,d , thus producing ≈100 neutron capture events/d in Borexino.

*Positioned Point Sources:* A series of point calibration sources is being designed which can be positioned throughout the detector volume. For high energy α's, a $^{222}$Rn-source will be used at a given position, supplementing the dispersed internal Rn radiations. We envision for the lower energy γ-rays a $^{7}$Be source (478 keV) and for pure β's a $^{32}$P source (1709 keV endpoint). For low energy α-particles, either $^{232}$Th (4.0 MeV) or $^{238}$U are considered. For line-β/γ sources $^{113}$Sn and $^{137}$Cs may be suitable.

*Calibration of the Neutrino Response:* In order to provide a direct demonstration of the overall ν response of the detector, calibration by means of a man-made sub-MeV ν-source with activities in the Megacurie range is foreseen. Such a final source test is important especially if the measured solar ν signal is null or close to the background level. Plans include both electron-neutrino ($^{51}$Cr) and electron-antineutrino ($^{90}$Sr) sources. A tunnel has been installed just below the external steel tank for inserting and retrieving the heavily shielded source containers on steel tracks.

### 4.3.3 Event Simulation

Neutrino-induced events and background have been studied by a simulation code, structured in three segments. In the first one[56] all the particles produced in a ν interaction or in radioactive decay are generated with a detailed modeling of the electromagnetic shower produced by the electrons or γ-rays. This leads to the energy-space characterization of these events. A second segment tracks the light emitted in the scintillator. It takes into account absorption, re-emission and scattering in the scintillator and in the buffer, as well as reflections on surfaces. A different version of this tracking has also been set up using the GEANT4 code.[57] In both codes the light collected on the PMT's is converted into time and charge signals. The third segment, the 'reconstruction code', performs the inverse process in which, starting from the PMT's signals, the space-time coordinates of the events are evaluated by maximum likelihood using a probability density function. The typical values obtained with this code for the (1σ) space and energy resolutions in the simulation of an electron of 1 MeV in the center of the IV are ≈8cm and ≈50 keV. These simulations have been successfully tested in the CTF data analysis.

## 4.4 External Background

The principal concern in Borexino is the non-shieldable internal background, due to the radioimpurities of the LS itself. Nevertheless, an important role in the detector design is also played by the need to suppress the γ background from external sources, such as the construction materials and the surrounding rocks, as well as the cosmic ray interactions in the underground environment. This impacts the architectural design and demands extraordinary care in the selection of materials of the detector by characterization of radiopurities. We briefly outline the methods used for material quality control, the on-line diagnostics available for determination of the overall external background, and some details on the effects of cosmic ray interactions.

### 4.4.1 Radiopurity of Detector Materials

The basic architecture of Borexino calls for high purity construction materials at every layer of the detector. Quality control of radiopurity specifications of every material, from the steel in the external tank to the scintillator in the IV as well as the materials of all liquid handling systems, has been performed by various counting methods. The major techniques were: Rn emanation measurements, Ge spectroscopy, mass spectroscopy and NAA, with progressively higher sensitivities but less wider applicability. Table 3 shows the Borexino design radiopurities compared with what has been achieved in CTF.

Samples of $N_2$, air and water were routinely assayed for Rn with Lucas cells[58] and later with miniature proportional counters. The key aspect of this technique is the concentration procedure of Rn with low blank values from large samples.[47,48,53] Rn emanation of detector components and materials is routinely tested at a sensitivity level of $50\mu Bq^{222}Rn$.[59] In related work, the permeation of Rn through various materials, especially the nylon film of the LS vessel was tested. This revealed major effects vital to the design of the IV.[60,49] Large volume Ge spectrometers with Rn free counting and special shielding assayed hundreds of samples (up to kg mass) for U/Th at a sensitivity of $\approx 10^{-10}$g/g. Most of the bulk material in the general construction were screened by this method, setting up an extensive database on bulk material radiopurity.[61,62] Inductively coupled mass spectrometry (ICPMS) is applicable to U,Th detection with a typical sensitivity of $10^{-12}$g/g, rising however to $10^{-15}$g/g for water in particular.[63] This technique has been applied to assay water samples, nylon material (after concentration by chemical ashing or digestion) and LS material (after concentration by water extraction). By far the most sensitive technique applicable to key impurity species, e.g. specific determination of $^{238}U$ in the $10^{-17}$g/g regime, NAA-ISAN was used to decide critical questions in the LS purity. Information on the purity of $^{40}K$ in PPO, the scintillation fluor, comes only from NAA. Besides U,Th and K, other long lived naturally occurring nuclides can also be measured with this technique.[41]

### 4.4.2 On-line Measurement of External Background

The major γ-ray and neutron background at the FV arises from the rock environment and the outer parts of the detector (OV steel sphere, PMT assemblies, the liquid buffer and IV walls). The radiopurity (as well as Rn emanation) of all these materials is known so that the γ-ray flux can be simulated precisely at any part of the detector.[56,61] The dominant source is the PMT array, especially the PMT glass, even with low activity glass. The 3.25m buffer liquid shields the FV from the PMT background. In this, 1.25 m of the shielding is active and the combined effect of passive shielding plus active rejection of external background energy deposition in the active region achieves 7-8 orders of magnitude in background reduction, to <0.1 event/d. The external background produces the largest fraction of events in the active buffer zone in the scintillator volume, thus, their magnitude and radial dependence can be used to separate the external component as well as to estimate the γ-ray leakage into the FV. Fig. 14 shows the result of a Monte-Carlo calculation of the external background for a lifetime of ≈10 days. It depicts the energy spectrum vs. the reconstructed radial distance of the event. No events are seen in the ν window in the FV. Background events begin to appear as the energy and the radial distance increases towards the edge of the FV, affecting, e.g., the window of the solar pep signal at (0.9-1.5) MeV. The problem is central for measuring the $^8$B-ν flux in the region of (3-6) MeV, measurable only in Borexino.

**Table 3**: Requirements on the radiopurity of detector materials for Borexino and values measured in the CTF

| Material | Borexino design | CTF achieved | Unit |
|---|---|---|---|
| Stainless steel | $\approx 10^{-9}$ | $\approx 10^{-9}$ | g/g of Th,U equiv. |
| External water | $\approx 10^{-10}$ | $\approx 10^{-14}$ | g/g of Th,U equiv. |
| PM | $\approx 10^{-8}$ | $\approx 10^{-8}$ | g/g of Th,U equiv. |
| Scintillator | $\approx 10^{-16}$ | $\approx 10^{-16}$ | g/g of Th,U equiv. |
| Scintillator | $\approx 10^{-18}$ | $\approx 10^{-18}$ | $^{14}C/^{12}C$ |

### 4.4.3 Cosmogenic Radioactivity

The effect of muons in Borexino is twofold:

*(i)* production of prompt events in the ν window; and

*(ii)* muon induced radioactivity by various reactions.[52,64]

The muon veto described above is the main defense against *(i)*. Veto effectiveness against *(ii)* depends basically on the lifetimes of the induced radioactivities. The dead time incurred by vetoes using a muon signal is limited to a few seconds. Thus, activities with lifetimes shorter than this range can be vetoed by the muon detector. The absence of isotopes heavier than $^{12,13}$C (and some $^{16,17,18}$O in the fluor) is a key factor that limits the list of possible activities. A fundamental datum is the CTF result of ≈0.3 neutrons/ton,d correlated with muons, that sets a limiting rate on all possible spallation reactions. This translates to some 100 spallations per day in the Borexino FV. The main question is what fraction of these events produce long-lived radioactivity that eludes the muon veto. To obtain more data on this point, the interaction of high energy muons on carbon and on liquid scintillators was studied in an experiment at the SPS muon beam at CERN.[64] Production cross-sections for possible radionuclides from the interaction of 100 and 190 GeV muons on $^{12}$C were measured.

The possible radioactivities from a C target too long-lived to be vetoed by the muon signal are $^{11}$Be, $^{11}$C, $^{10}$C and $^7$Be. $^{11}$Be is relatively harmless since its production cross-section is very small. $^{10,11}$C are both β$^+$ emitters so that their decay spectra are offset by 1.02 MeV out of the Be-ν signal window (0.25-0.8) MeV by calorimetric summing of the annihilation radiations with the β$^+$ spectrum.

The production of $^{10,11}$C from $^{12}$C releases neutrons which will be captured by protons after a typical delay of several 100μs to emit a taggable 2.2 MeV γ-ray. $^{10}$C with a mean lifetime of τ=28s could possibly be tagged reasonably near the located site of the n-capture. The case of $^{11}$C with τ=1765s may be less straightforward. $^7$Be presents direct interference since its 478 keV γ-ray produces background in the signal window itself. However, the event rate in the FV at saturation after nearly a year is expected to be only ≈0.4 counts/d.

**Figure 14**: Monte Carlo simulation of the external background, corresponding to 10 days of data acquisition: Plotted is the energy [MeV] vs. the distance from the center of the IV [cm]. The (empty) box at the lower left corresponds to the energy-position window of interest for $^7$Be-$\nu_e$ in Borexino (the innermost 2.25m are truncated). The shade code scale on the right is the $\log_{10}$ of the number of events.

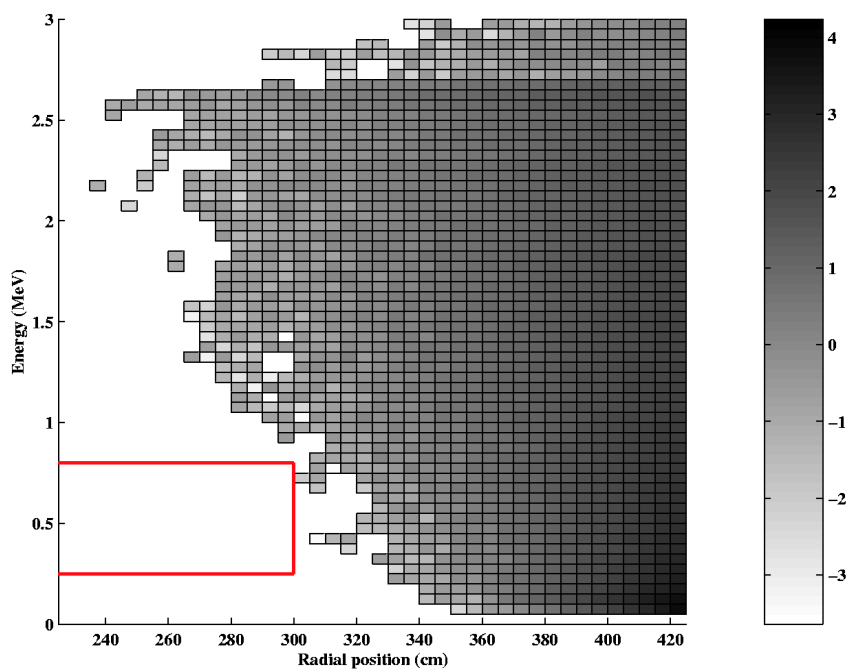

## 5. Response to Non-Standard Solar Neutrinos

As described in Sec. 2, the results of the present solar ν experiments are globally compatible with several ν conversion scenarios. On a quantitative basis set by the SSM fluxes, the possible ν parameters are restricted to a few islands in the $\Delta m^2$-$\sin^2(2\theta)$ space (see Figures 5,6). The urgent need is to find specific solar-model independent effects that can prove flavor conversion. Two such effects for the high energy (>6 MeV) $^8$B-ν signal - a deviation from the Fermi spectral shape and a day/night signal variation - have been sought in high precision SK data after a lifetime of >1000 days. Neither effect has been observed.[65,11] The remaining high energy test, a difference in the CC and NC signal rates, is being pursued by SNO.

Borexino will unveil a new real-time energy window to search for conversion effects, viz. the *low energy* solar ν spectrum focusing on the $^7$Be-neutrinos. With the evidence for conversion coming from the $^7$Be-$^8$B problem rather than from model-dependent flux deficits, strong effects at low energy are predicted, and thus a distinct signature in Borexino should be seen. The basic data expected from Borexino are the magnitude of the ν-e scattering signal rate due to mono-energetic $^7$Be solar ν's and a possible time dependence of this rate over day/night or seasonal periods. Also, the spectral shape of the $^8$B ν-e scattering signal in the lower energy (3-6) MeV regime may be measured.

$^7$Be Neutrino Signal Rates: With 100 tons of the PC scintillator target, the SSM and a standard ν predict, in the energy window (250-800) keV, a signal rate of 55/d, 80% of which is due to $^7$Be-neutrinos. Fig. 15 shows a simulated signal spectrum observable in Borexino and the background expected for $10^{-16}$gU/g. Based on the class of radiopurity and tag efficiencies demonstrated in the CTF, such radiopurity level is achievable in Borexino. If the solar $\nu_e$ is fully converted to other active flavors, the rate drops to a hard core minimum of ≈11/d, arising only from the NC interaction. With reasonable precision, this low signal would strongly suggest (if not prove) flavor conversion rather than an astrophysical effect since such a rate for $^7$Be electron neutrinos is incompatible with the observed $^8$B-ν signal rate. Evidence for non-standard ν's would become significantly stronger if the Borexino signal drops further, below the hard core value, in the event of full conversion to sterile neutrinos. Borexino is sensitive to this scenario if the background can be reduced significantly below that in Fig. 15. In that case, just an upper limit that rules out the hard-core signal rate is sufficient to prove conversion to sterile ν's unambiguously. For *intermediate* rates between the hard core and the SSM limits, inference of flavor conversion is less definitive.

The major conversion scenarios are the MSW effect and vacuum oscillations, which present four different parameter regimes as discussed in Sec. 2 and summarized in Figures 4-6. Figure 16 is a plot of iso-depression lines of the expected signal due to

**Figure 15**: Monte Carlo simulation of ν signals (in arbitrary units) and background. Plotted are the $^7$Be-ν signal expected from the SSM (dotted-dashed), the signal from all other neutrino sources together (dotted), the background (dashed), and the sum spectrum from all events (solid line). The background is calculated for $10^{-16}$ gU/g, an α/β discrimination of 90%, and 1σ statistical cuts.

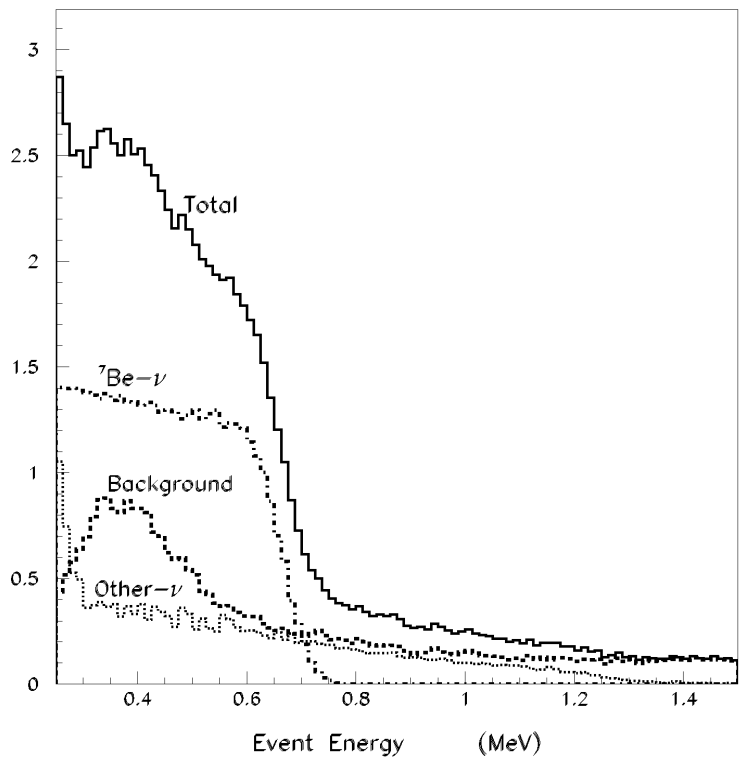

[7]Be neutrinos in the Borexino detector in the $\Delta m^2$-$\sin^2(2\theta)$ plane. As for the other neutrino types and for the absolute rates expected in Borexino, they are listed in Table 4 for the SSM and for the different MSW solutions. The MSW effect predicts significantly different rates for the small mixing (SMA) and the large mixing (LMA) regions. In the SMA, the [7]Be-ν's are nearly fully converted, leading to a signal rate close to the hard core value. Such a result would strongly suggest flavor conversion. In the LMA, the signal is typically reduced to $\approx$50% of the SSM value. In this case conclusions on flavor conversion can be reached only in conjunction with other experiments. In the LOW region, the rate is similar as for LMA, but a drastic day/night effect can be detected (see below).

*Temporal Variation of the [7]Be Neutrino Signal:* Unlike the rates, temporal variations of the signal are always decisive observables for flavor conversion. In two of the viable scenarios, ν's are reconverted after leaving the Sun either *(i)* by vacuum oscillations on the way to the Earth or *(ii)* by matter regeneration in passage through the Earth.    In both cases, the result is a time variation of the signal:

*(i)* The vacuum oscillation effect depends on the variation of the Earth-Sun distance in the Earth's eccentric orbit, thus the time variation is seasonal. This scenario is valid for maximal mixing and very small mass parameters, $\Delta m^2 \approx 10^{-9}$ to $10^{-11} (eV/c^2)^2$. Fig. 17 shows the daily signal in its seasonal dependence for two values of $\Delta m^2$ that are consistent with data from SK.[11]

*(ii)* The passage through the Earth occurs nightly, thus, MSW regeneration of the electron flavor in the Earth matter enhances the signal during the night and produces a day/night effect (DNE) of the signal. The magnitude of the DNE depends on the ν path length through the Earth, the penetration towards the core (with increasing density of the Earth matter) and for statistics, the length of night hours over the year. In addition, the regeneration and thus the DNE depends also on the geographical latitude of LNGS as well as on the tilt of the Earth axis relative to the Sun, i.e. the day of the year. The effect thus changes throughout the year in a pattern characteristic of the geographical and of the ν parameters.

Earth regeneration effects are particularly enhanced in the LOW region. Borexino is well positioned to explore this because the largest variation occurs just at the energy of [7]Be-ν's. This can be seen in Fig. 4 which displays the DNE for different MSW scenarios. The mono-energetic feature of the [7]Be-line is particularly suitable for testing because the DNE is not fragmented over a wide energy range.[17,66]

*Low Energy Spectral Shape of [8]B Neutrinos:* A general characteristic of the energy dependence of flavor conversion by the MSW effect is a gradual decrease in flavor survival - and stronger observable effects - towards low energies. The distortion of the spectral shape of the [8]B signal (relative to that from a Fermi shape of the ν spectrum) thus increases towards the low end of the spectrum. The high energy part of the spectrum >6 MeV has been measured in water Čerenkov detectors Kamiokande, SK and now in SNO. In this region, so far, spectral deviations appear small, if any. It is therefore important to extend the data to lower energies[49], the only foreseeable chance for which lies in Borexino because of its low energy sensitivity although the relatively small target mass compared to SK is an inherent limitation. The expected rates for [8]B-ν's in Borexino for the different scenarios are included in Table 4.

**Table 4:** Solar neutrino counting rates expected via ν-e scattering per day in Borexino in three windows of the recoil electron energy for 4 scenarios:

-SSM, the standard solar model

-LMA, the large mixing angle solution

-SMA, the small mixing angle solution,

-LOW, the region of a day/night effect.

The rates are calculated for the characteristic parameters quoted in the caption of Fig. 5, for the mean Sun-Earth distance, for $113m^3$ fiducial volume, and for PC as scintillator. No radiative corrections are included. The case of vacuum oscillations is not shown in the Table. In this case the rate would be subject to strong seasonal variation. Hence, real-time detection of the mono-energetic $^7$Be-ν's enables a unique test of vacuum oscillations.

| 0.25 - 0.80 MeV | p-p | 0.22 | 0.15 | 0.08 | 0.13 |
|---|---|---|---|---|---|
| | $^7$Be | 43.3 | 24.4 | 9.20 | 22.8 |
| | p-e-p | 2.0 | 0.95 | 0.39 | 1.03 |
| | $^{13}$N | 4.0 | 2.27 | 0.87 | 2.13 |
| | $^{15}$O | 5.5 | 2.86 | 1.12 | 2.86 |
| | $^{17}$F | 0.07 | 0.03 | 0.01 | 0.03 |
| | $^8$B | 0.08 | 0.03 | 0.04 | 0.04 |
| | Sum | 55.2 | 30.7 | 11.7 | 29.0 |
| 0.80 - 1.50 MeV | p-e-p | 1.43 | 0.68 | 0.28 | 0.73 |
| | $^{13}$N | 0.13 | 0.07 | 0.03 | 0.07 |
| | $^{15}$O | 1.80 | 0.86 | 0.35 | 0.92 |
| | $^{17}$F | 0.02 | 0.01 | 0.00 | 0.01 |
| | $^8$B | 0.10 | 0.04 | 0.05 | 0.05 |
| | Sum | 3.48 | 1.66 | 0.71 | 1.78 |
| 1.50 - 5.50 MeV | $^8$B | 0.45 | 0.17 | 0.22 | 0.23 |

**Figure 16**: Iso-depression lines in the $\Delta m^2$-$\sin^2(2\theta)$ plane for the expected measurable signal due to [7]Be neutrinos in the Borexino detector. Plotted are contours for 25, 35, 50, 70 and 90 percent of the full SSM signal due to 862 keV [7]Be-ν's, respectively. The NC contribution from $\nu_\mu$ - e[-] scattering is included. Averaged over day and night and over the year. For absolute rates, see Table 4.

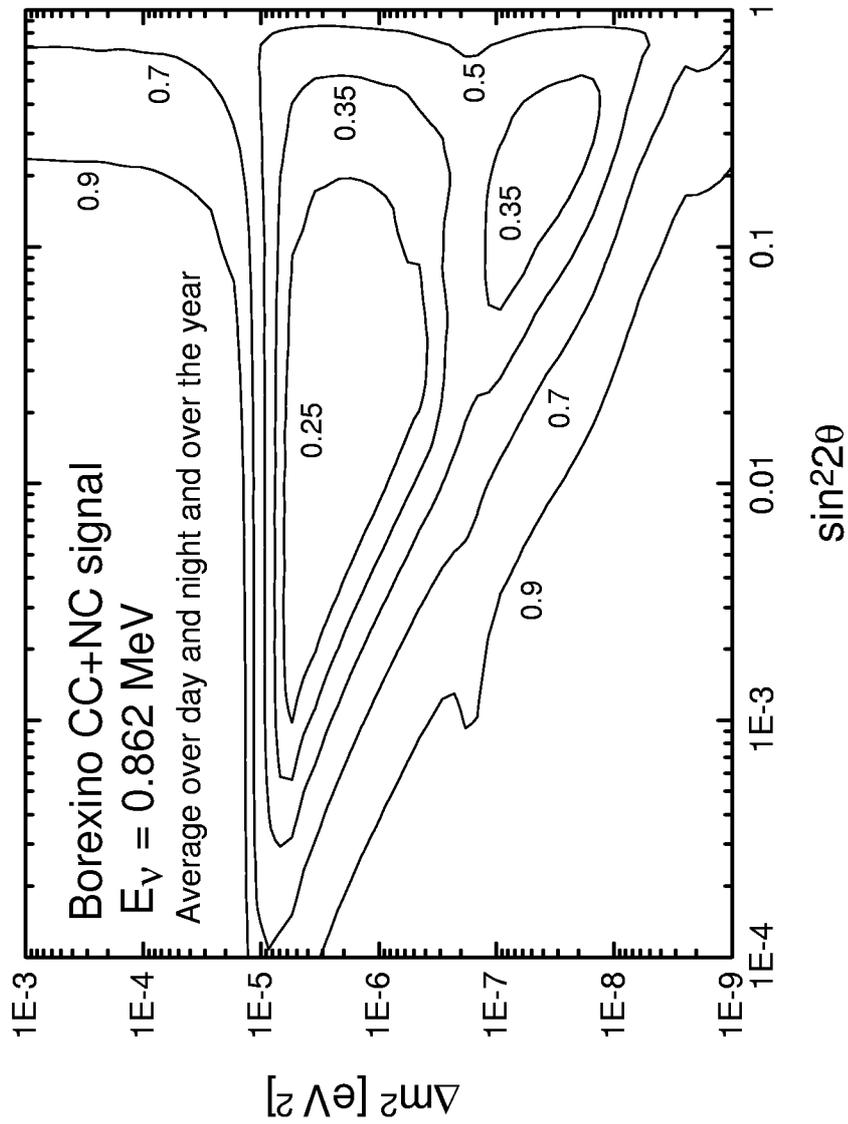

Borexino CC+NC signal
$E_\nu = 0.862$ MeV
Average over day and night and over the year

**Figure 17**: Annual variation of the daily $^7$Be-$\nu$ produced Borexino counting rate due to vacuum oscillations in combination with the eccentricity of the Earth's orbit. The dashed upper curve shows the no-oscillation signal with its purely geometrical variation ~$R^{-2}$ ($\approx 7\%$). Neutrino oscillations have particularly distinct effects on the mono-energetic $^7$Be-line. Two oscillation mass parameters have been selected to illustrate how accurately $\Delta m^2$ can be determined:

*(i)* solid line, $\Delta m^2 = 4.2 \times 10^{-10} (\text{eV/c}^2)^2$. This is the best fit for the Superkamiokande data.

*(ii)* dashed line, $\Delta m^2 = 3.2 \times 10^{-10} (\text{eV/c}^2)^2$, only marginally lower. Full mixing was assumed in both cases. The broadening of the $^7$Be-line due to the temperature in the solar center as well as the radial $^7$Be distribution in the solar core have been taken into account.

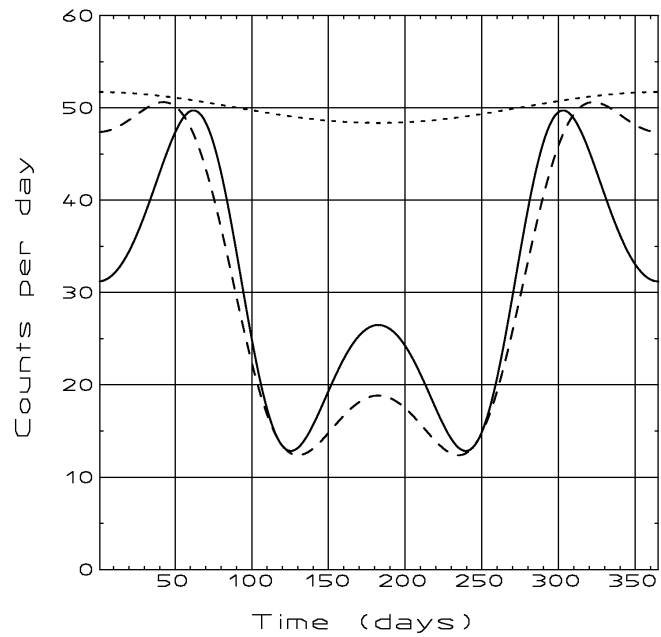

## 6. Non-solar Neutrino Science

While the basic objective of Borexino is the direct observation and measurement of the solar $^7$Be-ν flux, the facility can be applied to a broad range of frontier questions in particle physics, astrophysics and geophysics.[67] The unique low energy sensitivity and ultra-low background in Borexino bring new capabilities to attack problems in these fields. Much of this research can be undertaken simultaneously with solar ν observations. In particular, antineutrino ($\bar{\nu}_e$) spectroscopy can be performed simultaneously with a distinct tag independent of solar-ν spectroscopy. Other research needs additional facilities such as artificial ν sources or foreign targets in the scintillator. As in the case of solar ν's, the CTF is well suited for testing the feasibility of novel methods to attack some of the questions.

### *6.1 Antineutrino Science*

*Detection of Low $\bar{\nu}_e$ fluxes:* By far the best method to detect $\bar{\nu}_e$ is the classic Reines reaction of capture by protons in the scintillator liquid : $\bar{\nu}_e + p \rightarrow n + e^+$. The positron signal energy (kinetic energy + 1.02 MeV annihilation energy) E = E($\bar{\nu}_e$) − Q where the threshold energy Q = 1.8 MeV. Thus, even at threshold E($\bar{\nu}_e$)=Q, this reaction produces an easily detectable signal at 1.02 MeV. The $\bar{\nu}_e$ tag is made possible by a delayed coincidence of the positron and by a 2.2 MeV γ-ray emitted by capture of the neutron on a proton after a delay of ≈200μs caused by the moderation of the neutron to thermal energies. The tag suppresses background by a factor ≈100, thus, the active buffer is not crucial. In favorable circumstances, the entire scintillator mass of 300 tons may be utilized. One of the few sources of correlated background is muon induced activities that emit β-neutron cascades. However, all such cases have lifetimes τ<1s. Thus they can be vetoed by the muon signal. Overall, a signal rate as low as ≈1 $\bar{\nu}_e$ event per year and 300 tons appears measurable. The interesting sources of $\bar{\nu}_e$ are supernovae, the Sun, the Earth and nuclear power reactors, which can be distinguished from each other mainly by the characteristically different energy spectra of the signals. The sensitivity limit (1 event/yr) corresponds to a flux of $5 \times 10^2$ $\bar{\nu}_e$/cm$^2$,s or ≈$10^{-4} \times \varphi_\nu$($^8$B). Borexino is thus one of the most sensitive $\bar{\nu}_e$ detectors ever designed and it is tailored to several problems in particle physics, geophysics and astrophysics, all of which predict low $\bar{\nu}_e$ fluxes.

*Solar Antineutrinos:* The SSM predicts no $\bar{\nu}_e$ emission from the Sun. Non-standard ν physics however, can create $\bar{\nu}_e$ by some conversion mechanisms. One such model, still viable, is based on an off-diagonal or 'transition' ν magnetic moment ($\mu_{tr}$) allowed theoretically for a *Majorana* neutrino. The best present limit on $\mu_{tr}$ , based on astrophysical grounds, is $3 \times 10^{-12}$ $\mu_B$ (Bohr magneton).[68] Laboratory limits are less stringent by typically 2 orders of magnitude.[68] The interaction of $\mu_{tr}$ with solar magnetic fields may produce a spin-flavor precession, i.e. a spin-flip and a flavor conversion resulting in μ or τ antineutrinos.[69] The optimal conditions apply to $^8$B ν's in the LOW mass area of the MSW map, i.e. $\Delta m^2 \approx 10^{-7}$(eV/c$^2$)$^2$ and near-maximal mixing. So far, this effect causes only solar ν$_e$ disappearance. However, an interesting consequence is the *appearance* of solar *electron* antineutrinos at the Earth. Because of the maximal mixing inherent in the model, the resulting μ/τ antineutrinos can convert to $\bar{\nu}_e$ simply by vacuum oscillation.[17,66,70] The magnitude of this $\bar{\nu}_e$ flux depends on M$^2$ with M = $\mu_{tr} \times$B (B is the solar magnetic field). The current experimental limit[71] on the flux of $\bar{\nu}_e$ with energy >7 MeV is <$5\times10^4$/cm$^2$, which corresponds to an event rate of <110 $\bar{\nu}_e$/yr in Borexino. This already sets a limit of M<$7\times10^{-11}\mu_B$ kGauss. At a sensitivity of 1 event/yr, Borexino can reach a limit M <$10^{-12}\mu_B$ kGauss.

*Antineutrinos from the Earth's Interior:* A key factor in the conceptual foundations of geophysical models of the Earth's interior is the radiogenic heat from the decay of U and Th in the Earth's crust, currently believed to be ≈40% of the total heat flow of ≈40TW. Borexino can make a basic contribution to testing these models by detecting $\bar{\nu}_e$ emitted by these nuclides in the range of (1.8-3.3) MeV. Three specific aspects make such a measurement valuable[72]:

*(i)* One obtains a 'whole Earth' view of the crustal U/Th abundance instead of sampling by laborious field work;

*(ii)* The U and Th abundances can be separately measured;

*(iii)* by combining data from Borexino (Eurasian plate) with those from the KAMLAND detector of similar design planned in Japan (interface of Asian and oceanic crusts), the relative distribution of U/Th in the continental and oceanic crusts may be probed. Depending on the geophysical model, $\bar{\nu}_e$ rates between 10 and 30 events/yr can be expected in Borexino.[72]

*Long-baseline $\bar{\nu}_e$ from European Reactors*: The flux of $\bar{\nu}_e$ from astrophysical sources (including the Earth) are essentially model-dependent. Nuclear power reactors emit $\bar{\nu}_e$ with a known flux and spectral shape with energies up to ≈8 MeV. Borexino is sensitive to this flux from power reactors situated all over Europe at an average baseline distance of ≈750km. There are no reactors in Italy itself. The multi-reactor $\bar{\nu}_e$ flux produces an accurately (≈5%) predictable signal in Borexino of ≈30 events/yr. The well defined set-up offered by this combination makes an ideal terrestrial long-baseline experiment for a model-free search for vacuum ν oscillations.[73]

The sensitivity for ν oscillations is in the range $\Delta m^2 \approx [10^{-3}\text{-}10^{-5}](\text{eV/c}^2)^2$, filling the gap between the results of the CHOOZ experiment and solar neutrinos. The European reactor-Borexino combination is thus also suitable to probe ν oscillations in the LMA-MSW area, as it is currently planned for the Kamland detector.

### 6.2 Neutrinos and Antineutrinos from Type-II Supernovae

The occurrence of a supernova burst creates a ν flux of all types in the energy range of 10's of MeV. They occur as a short pulse that lasts for several seconds. Two factors make its detection in Borexino interesting.[74] As the first major detector with liquid scintillator operating at low energies, Borexino may be unique despite the relatively low detector mass because:

*(i)* low energies in the Fermi-Dirac spectrum of the supernova ν's can be probed;

*(ii)* $\bar{\nu}_e$ can be detected with high sensitivity and

*(iii)* Supernova neutrino interactions with $^{12}$C in the liquid scintillator offer a unique tool for a CC/NC analysis of the ν flux by combining the results from the reactions $^{12}$C$(\nu_e,e^-)^{12}$N, $^{12}$C$(\bar{\nu}_e,e^+)^{12}$B and $^{12}$C$(\nu_x,\nu_x)^{12}$C*(15.1 MeV).

The signals are estimated to be ≈80 $\bar{\nu}_e$ - events and ≈30 events from reactions on $^{12}$C for a $3\times10^{53}$erg burst at a distance of 10kpc. The principal results of interest to the supernova mechanism as well as to ν physics are:

*(i)* the fluxes of different flavors via the NC/CC data;
*(ii)* the $\bar{\nu}_e$ spectrum;
*(iii)* limits on the $\nu$ mass via the difference in times of arrival between the CC and NC signals; and
*(iv)* possible neutrino oscillation signals.

### 6.3 Neutrino Physics with Megacurie Sources

Man-made sources of neutrinos from intense radioactive $^{51}$Cr sources have been used in the Ga experiments to subject the radiochemical detectors to rigorous throughput tests of all procedures and components.[75] Gallex has expanded this approach to achieve a *real calibration* of the detector by complementing the source experiment with a high-statistics $^{71}$As doping experiment.[76] The application of a Megacurie neutrino source is also foreseen for Borexino to demonstrate the performance of the detector. This is especially valuable in the event of a very small or null signal (e.g. with complete conversion to active or sterile neutrinos). In addition, as a detector based on $\nu$-e scattering, Borexino offers the interesting opportunity to probe the scattering mechanism to the level of precision sufficient to uncover new physics. The recoil electron profile in $\nu$-e scattering deviates at low energies from the normal weak-interaction mediated scattering if the neutrino carries a static magnetic moment $\mu_\nu$. As a massive detector with uniquely low background and low energy sensitivity, Borexino offers a new means for searching for $\mu_\nu$ using artificial neutrino sources[17,77] such as $^{90}$Sr-$^{90}$Y (which emits *antineutrinos*) and $^{51}$Cr which emits neutrinos. These ideas have recently been updated[78] for the geometry of Borexino with consideration of the non-removable background of solar neutrino scattering for various $\nu$ scenarios. In the case of a $^{51}$Cr $\nu$ source, the isotope enriched source material is potentially available and it offers the advantage of a mono-energetic line at lower energies than $^{90}$Sr, although the shorter lifetime is a drawback in this application. The sensitivity on $\mu_\nu$ attainable in Borexino[17] for a 1 MCi source is $\mu_\nu \approx 5 \times 10^{-11} \mu_B$ for $^{51}$Cr and $\approx 3 \times 10^{-11} \mu_B$ for $^{90}$Sr. The latter sensitivity could be enhanced by the technical feasibility of multi-MCi sources. A tunnel facility for inserting and withdrawing sources under the Borexino detector is already built into the detector.

### 6.4 Double Beta Decay in $^{136}$Xe

The phenomenon of $\beta\beta$-decay offers the only means yet known for identifying the Majorana nature of the neutrino. The search for $\beta\beta$-decay has been traditionally carried out with target masses in the tens of kg scale. Borexino offers an opportunity for enhancing this search on the multi-ton scale[17,79] by exploiting the high solubility of *noble* gases in liquid scintillators without affecting radiopurity or scintillator efficiency. A favorable candidate is $^{136}$Xe. Tests could be planned first in the CTFII on the scale of 10kg of enriched $^{136}$Xe for assessing signal quality, background etc. If successful, the measurements could be extended in Borexino itself with $\approx 2$ tons of $^{136}$Xe.

## 7. Conclusions and Outlook

The 'Solar Neutrino Problem' remains unsolved but restated more sharply than ever in terms of the $^7$Be-$^8$B problem. Apparent paradoxes that cut through caveats with compelling simplicity have arisen before in weak interaction physics and led to scientific revolutions. An apparent violation of energy conservation led to the postulation of the neutrino itself. The $\tau$-$\theta$ puzzle led to parity violation and a helical neutrino. The $^7$Be-$^8$B problem signals another breakthrough, this time towards a flavor mixed, massive neutrino, with emphasis on $\nu_e$-$\nu_\mu$ mixing. The same phenomenon, neutrino oscillations, is strongly indicated by the SK observation of $\nu_\mu$-deficits relative to $\nu_e$ in cosmic ray produced atmospheric neutrinos that have penetrated the Earth[80], indicative of $\nu_\mu$-$\nu_\tau$ mixing. How will these developments take shape in the next few years and what role will Borexino play?

Pointers to the immediate future can be gauged from the ongoing accumulation of SK solar neutrino data. The experiment has so far produced no 'smoking gun' for solar neutrino conversion despite the admirable precision of its data. Limits on spectral deviations in the $^8$B spectrum and the day/night effect continue to point to a lack of evidence for any of the conversion scenarios, MSW or vacuum oscillations. In the next step the newly started SNO will reexamine the $^8$B spectral shape for small distortions, with higher sensitivity via inverse $\beta$-decay. The ratio of NC/CC signals in SNO can establish conversion in almost all scenarios, but cannot give any decisive result if conversion occurs to sterile neutrinos.

The uncertainties in the astrophysical models and in the nuclear cross-section of $^7$Be+p do not rule out smaller SSM $^8$B-$\nu$ fluxes, smaller deficits thus, smaller conversion effects at $^8$B energies to start with. This scenario is in fact consistent with the present SK results. On the other hand, the very existence of the $^7$Be-$^8$B flux paradox is a strong indication of flavor conversion in the sub-MeV energy range. Borexino will provide the first opportunity to explore this regime with two kinds of data:

*(i)* the value of the $^7$Be-$\nu$ flux, and

*(ii)* 'appearance' effects such as time variation of the solar signal.

In scenarios where flavor survival varies slowly with energy (such as SMA, LMA), the Borexino signal is time independent (besides the trivial seasonal 7% $R^{-2}$ variation) and is a measure of the $^7$Be-$\nu$ flux that depends, in principle, on the solar model. However, if the measured signal is decidedly at or below the NC 'hard core' limit, such a result would be a strong indication for conversion to sterile neutrinos.

The vacuum oscillation scenario predicts a rapid variation of the survival probability with the energy and with the distance from the Sun. This effect is hardly detectable in SK and in SNO, hampered by an average over the wide range of energies in the $^8$B-$\nu$ spectrum. Borexino, on the other hand, looks for a mono-energetic neutrino signal and it can thus decouple the energy dependence from the periodical change of the Sun-Earth distance. Borexino has a unique ability to probe the vacuum oscillation scenario, through the detection of a time variation in the signal that is significantly larger than the $R^{-2}$ effect.

The time scale of the variations range from a seasonal to monthly and even daily basis for vacuum oscillations with $\Delta m^2 \approx 10^{-11}$ up to $\approx 10^{-8} (eV/c^2)^2$. At somewhat higher $\Delta m^2$ values, $\approx 10^{-7} (eV/c^2)^2$, the LOW mass MSW scenario produces a day/night effect. Such time variations are 'smoking guns' for flavor conversion regardless whether neutrinos convert to active or sterile species. In either of the above two possibilities, Borexino would play a decisive and indispensable role in demonstrating neutrino flavor conversion, regardless of the type of non-standard neutrino that generates it in nature.

## Acknowledgements

Many individuals besides the authors have contributed to the development of the Borexino project. We wish to acknowledge particularly the following who made significant contributions in early stages: R. Cereseto, N. Darnton, A. Falgiani, T. Goldbrunner, J. Jochum, M. Johnson, A. Manco, A. Nostro, S. Pakvasa, M. Parodi, A. Perotti, G. Pieri, A. Preda, P. Raghavan, P. Rothschild, P. Ullucci and members of IRMM for the measurements made in Geel. The collaboration wants to thank the Laboratori Nazionali del Gran Sasso [LNGS] and A. Bettini for continuous help and support. We sincerely thank the funding bodies: Istituto Nazionale di Fisica Nucleare [INFN], and Ministero della Universita' e della Ricerca Scientifica e Tecnologica [MURST] (Italy); IN2P3 (France); Bundesministerium für Bildung, Wissenschaft, Forschung und Technologie [BMBF], Deutsche Forschungsgemeinschaft [DFG] and Max-Planck- Gesellschaft [MPG] (Germany); the National Science Foundation and Bell Laboratories (USA); the Natural Sciences and Engineering Research Council of Canada, and agencies in Hungary, Poland and Russia for their generous support of this project.